\documentclass[12pt]{article}
\pdfoutput=1

\usepackage{dsfont}
\usepackage{epsfig}
\usepackage{psfrag}
\usepackage{latexsym}
\usepackage[DIV13]{typearea}
\usepackage{amsmath}
\usepackage{amssymb}
\usepackage{amsfonts}
\usepackage{mathrsfs}
\usepackage{indentfirst}
\usepackage{cite}
\usepackage{bbold}
\usepackage[footnotesize]{caption}
\usepackage{graphicx}
\usepackage[center,footnotesize,hang]{subfigure}
\usepackage{url}
\usepackage{verbatim}
\usepackage{color}
\usepackage{inputenc}
\usepackage{slashed}

\newcommand{\la}{\lambda}
\newcommand{\La}{\Lambda_f}

\newcommand{\LL}{\mathscr{L}}

 \def\cO{{\mathcal O}}

\newcommand{\hc}{\text{h.c.}}
\newcommand{\unity}{\mathbb{1}}

\newcommand{\nn}{\nonumber}
\newcommand{\mean}[1]{\langle#1\rangle}
\newcommand{\derp}{\partial}
\newcommand{\ov}[1]{\overline{#1}}
\newcommand{\tr}{\text{Tr}}

\newcommand{\beq}{\begin{equation}}
\newcommand{\eeq}{\end{equation}}
\newcommand{\bac}{\beq\begin{array}}
\newcommand{\eac}{\end{array}\eeq}
\newcommand{\ba}{\begin{array}}
\newcommand{\ea}{\end{array}}
\newcommand{\bea}{\begin{eqnarray}}
\newcommand{\eea}{\end{eqnarray}}

\newcommand{\appendixA}{\setcounter{equation}{0}
\def\theequation{\rm{A}.\arabic{equation}}\section*}
\newcommand{\appendixB}{\setcounter{equation}{0}
\def\theequation{\rm{B}.\arabic{equation}}\section*}
\newcommand{\appendixC}{\setcounter{equation}{0}
\def\theequation{\rm{C}.\arabic{equation}}\section*}

\begin{document}
\begin{titlepage}
\vspace*{-1cm}
\phantom{hep-ph/***}

\flushright
\hfil{FTUAM-11-39}\\
\hfil{IFT-UAM/CSIC-11-09}\\
\hfil{TUM-HEP-796/11}\\
\hfil{DFPD-11-TH-2}

\vskip 1.5cm
\begin{center}
{\Large\bf On The Scalar Potential of Minimal Flavour Violation}
\vskip .3cm
\end{center}
\vskip 0.5  cm
\begin{center}
{\footnotesize
{\large R. Alonso}~$^{a)}$\footnote{e-mail address: rod.alonso@estudiante.uam.es},
{\large M.B. Gavela}~$^{a)}$\footnote{e-mail address: belen.gavela@uam.es},
{\large L. Merlo}~$^{b,c)}$\footnote{e-mail address: luca.merlo@ph.tum.de} {\large and}
{\large S. Rigolin}~$^{d)}$\footnote{e-mail address: stefano.rigolin@pd.infn.it}
\\
\vskip .4cm
$^{a)}$~Departamento de F\'isica Te\'orica, Universidad Aut\'onoma de Madrid and
\\
Instituto de F\'{\i}sica Te\'orica IFT-UAM/CSIC, Cantoblanco, 28049 Madrid, Spain
\\
\vskip .1cm
$^{b)}$~Physik-Department, Technische Universit\"at M\"unchen, 
\\
James-Franck-Strasse, D-85748 Garching, Germany
\\
\vskip .1cm
$^{c)}$~
TUM Institute for Advanced Study, Technische Universit\"at M\"unchen, \\
Lichtenbergstrasse 2a, D-85748 Garching, Germany
\\
\vskip .1cm
$^{d)}$~Dipartimento di Fisica Galileo Galilei, Universit\`a di Padova and \\
INFN, Sezione di Padova, Via Marzolo~8, I-35131 Padua, Italy
}
\end{center}
\vskip 3cm
\begin{abstract}
\noindent

Assuming the Minimal Flavour Violation hypothesis, we derive the general scalar potential for fields whose background 
values are the Yukawa couplings. We analyze the minimum of the potential and discuss the fine-tuning required to 
dynamically generate the mass hierarchies and the mixings between different quark generations. Two main cases are 
considered, corresponding to Yukawa interactions being effective operators of dimension five or six (or, equivalently, 
resulting from bi-fundamental and fundamental scalar fields, respectively). 
At the renormalizable and classical level, no mixing is naturally induced from dimension five Yukawa operators. 
On the contrary, from dimension six Yukawa operators one mixing angle and a strong mass hierarchy among the 
generations result.
\end{abstract}
\end{titlepage}
\setcounter{footnote}{0}

%
%

\section{Introduction}

After years of intense searches, all flavour processes observed in the hadronic sector, from rare decays 
measurements in the kaon and pion sectors to superB--factories results, are well in agreement with the 
expectations of the Standard Model of particle physics (SM). To say that all flavour processes are consistent 
with the SM predictions is tantamount to state that all flavour effects observed until now are consistent with 
being generated through the Yukawa couplings, which are the sole vehicles of flavour and CP violation in the SM. 
  
Nevertheless, the origin of fermion masses and mixings remains the most unsatisfactory question in the visible 
sector of nature: it involves important fine-tunings and lack of predictivity, as essentially for each mass or 
mixing angle a new parameter is added by hand to the SM. It is commonly expected that an underlying dynamics will 
provide a rationale for the observed patterns.

The hypothesis of {\it Minimal Flavour Violation} (MFV)~\cite{DGIS:MFV} is a humble, matter-of-fact, and highly 
predictive working frame built only on: i) the assumption that, at low energies, the Yukawa couplings are the 
only sources of flavour and CP violation both in the SM and in whatever may be the flavour theory beyond it, 
abiding in this way to the experimental indications mentioned above; ii) the use of the flavour symmetries which 
the SM exhibits in the limit of vanishing Yukawa couplings.
 
Indeed, the hadronic part of the SM Lagrangian, in the absence of quark Yukawa terms, exhibits a flavour symmetry given by
\beq
G_f=SU(3)_{Q_L}\times SU(3)_{U_R}\times SU(3)_{D_R}\,,
\label{3famsym}
\eeq
plus three extra $U(1)$ factors corresponding to the baryon number, the hypercharge and the Peccei-Quinn 
symmetry\cite{PQ:U1PC}.  The non-abelian subgroup $G_f$  controls the flavour structure of the Yukawa matrices, 
and we focus on it for the remainder of this paper. Under $G_f$, the $SU(2)_L$ quark 
doublet, $Q_L$, and the $SU(2)_L$ quark singlets, $U_R$ and $D_R$, transform as: 
\beq
Q_L\sim(3,1,1)\,,\qquad\qquad
U_R\sim(1,3,1)\,,\qquad\qquad
D_R\sim(1,1,3)\;.
\label{QuarksTrans}
\eeq
The SM Yukawa interactions break explicitly the flavour symmetry: 
\beq
\LL_Y=\ov{Q}_LY_DD_RH+\ov{Q}_LY_UU_R\tilde{H}+\hc
\eeq
The technical realization of the MFV ansatz promotes the Yukawa couplings $Y_{U,D}$ to be spurion fields 
which transform under $G_f$ as
\beq
Y_U\sim(3,\ov{3},1)\;,\qquad\qquad 
Y_D\sim(3,1,\ov{3})\;,
\label{spurions}
\eeq
recovering the invariance under $G_f$ of the full SM Lagrangian. Following the usual 
MFV convention for the Yukawas, one defines
\beq
Y_D=\left(
        \begin{array}{ccc}
           y_d  & 0 & 0 \\
            0 & y_s & 0 \\
            0 & 0 &  y_b \\
        \end{array}
\right)\;,\qquad\qquad
Y_U={\mathcal V}^\dag_{CKM}\left(
        \begin{array}{ccc}
           y_u  & 0 & 0 \\
            0 & y_c & 0 \\
            0 & 0 &  y_t \\
        \end{array}\right)\;,
\label{SpurionsVEVs}
\eeq
with ${\mathcal V}_{CKM}$ being the usual quark mixing matrix, encoding three angles and one CP-odd phase. 

MFV is not a model of flavour and the value of the new dynamical flavour scale $\Lambda_f$ is not fixed: 
it does not determine the energy scale at which new flavour effects will show up. Nevertheless it is quite  
successful in predicting precise and constrained relations between different flavour transitions, to be observed 
whenever the new physics scale becomes experimentally accessible~\cite{Buras:MFVandBeyond}. The reason is that in the 
MFV framework the coefficients of all SM--gauge invariant operators have a fixed flavour structure in terms of Yukawa couplings,  so as to make the operator invariant under $G_f$, plus the fact that the top Yukawa coupling may dominate any coefficient in which it participates~\footnote{This is modified, though, in some MFV versions 
such as two--Higgs doublet models~\cite{DGIS:MFV} with extra discrete symmetries~\cite{Branco}, or in models with strong 
dynamics~\cite{Kagan}}.

MFV sheds also an interesting light on the relative size of the electroweak and the flavour scale. The origin of all 
visible masses and the family structure are the two major unresolved puzzles of the SM and it is unknown whether 
a relation exists between the nature and  size of those two scales. While the electroweak data, and the theoretical 
fine-tunings they require, suggest that new physics should appear around the TeV scale, traditional model-independent 
limits on the flavour scale $\Lambda_f$ point to order(s) of magnitude larger values~\cite{INP:GenericFS}. Within MFV 
both sizes could be reconciled instead around the TeV scale, due to the Yukawa suppression of the flavour-changing 
operator coefficients. This holds either assuming only the SM as the 
renormalizable theory~\cite{DGIS:MFV} or in beyond the SM scenarios (BSM), such as supersymmetric~\cite{LPR:MFVsusy} 
or extradimensional~\cite{FPR:MFVextraD} versions of the MFV ansatz\footnote{The BSM theory may introduce more 
than one distinct flavour scale: 
this work sticks to a conservative and minimalist approach, focusing on the physics 
related to $\Lambda_f$ as described above.}.

It is unlikely that MFV holds at all scales~\cite{Isidori2-bis}. MFV assumes a new dynamical scale $\Lambda_f$, 
which points to MFV being just an accidental low-energy property of the theory. In this sense, MFV implicitly 
points to a dynamical origin for the values of the Yukawa couplings. The latter may correspond to the vacuum 
expectation values (vevs) of elementary or composite fields or combinations of them. In other words, 
the spurions may be promoted to fields, usually called flavons. For instance, in the first formulation of MFV by 
Chivukula and Georgi~\cite{CG:MFV}, the Yukawa couplings corresponded to a fermion condensate. In this work, we 
further explore the dynamical character of the flavons, in a rather model--independent way. 
 
\begin{figure}[h!]
\centering
\includegraphics[height =5cm]{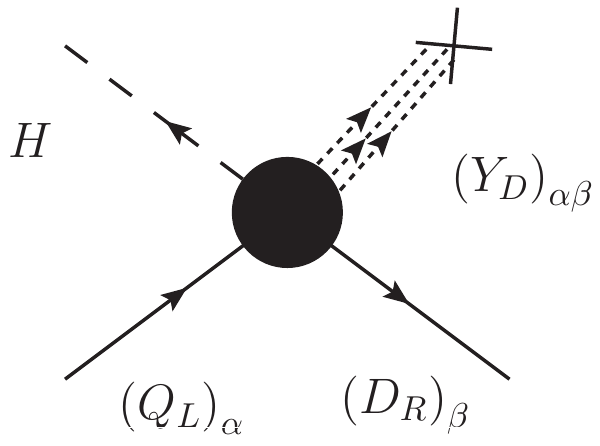}
\caption{Effective Yukawa coupling.}
\label{EffectiveYukawa}
\end{figure}

The Yukawa interactions may be then seen as effective operators of dimension larger than four 
--denominated {\it Yukawa operators} in what follows-- weighted down by powers of the large flavour 
scale\footnote{For instance, a possible realization among many takes $\Lambda_f$ to be the mass 
of heavy flavour mediators in some BSM theory \cite{Syms}: at energies $E<\Lambda_f$, they can be integrated out resulting 
in $d>4$ operators involving the SM fields and the flavons.} $\Lambda_f$. The precise dimension $d$ of the 
Yukawa operators is not determined, as illustrated in Fig.~\ref{EffectiveYukawa}. As long as the vev to be 
taken by the flavon fields is smaller than $\Lambda_f$, an analysis ordered by inverse powers of this scale 
is a sensible approach. 
The simplest case is that of a $d=5$ operator: 
\beq
\LL_Y=\ov{Q}_L\dfrac{\Sigma_d}{\Lambda_f}D_RH+\ov{Q}_L\dfrac{\Sigma_u}{\Lambda_f}U_R\tilde{H}+\hc\;,
\label{LagrangianBi_Fund}
\eeq
with the scalar flavons $\Sigma_d$ and $\Sigma_u$ being dynamical fields in the bi--fundamental representation of $G_f$ (i.e. 
$\Sigma_u\sim(3,\ov{3},1)$ and $\Sigma_d\sim(3,1,\ov{3}$), see eq.~(\ref{spurions})) such that\footnote{The 
Goldstone bosons that would result from the spontaneous breaking of a continuous global flavour symmetry, 
may be avoided for instance by gauging the symmetry. In practical realizations, this in turn tends to induce 
dangerous flavour--changing neutral currents mediated by the new gauge bosons. A new promising avenues to cope 
with this problem has been recently proposed in ref.~\cite{GRV:SU3gauged,Feldmann:SU5gauged}.}
\beq
Y_D \equiv \dfrac{\mean{\Sigma_d}}{\Lambda_f}\,, \qquad \qquad
Y_U \equiv \dfrac{\mean{\Sigma_u}}{\Lambda_f}\,.
\label{dim5Y}
\eeq

An alternative realization, that we also explore below, is that of a $d=6$ Yukawa operator, involving generically 
two scalar flavons for each spurion,
 \bea
\LL_Y=\ov{Q}_L\frac{\chi_d^L\chi_d^{R\dag}}{\Lambda_f^2}D_RH+\ov{Q}_L\frac{\chi_u^L\chi_u^{R\dag}}
{\Lambda_f^2}U_R\tilde{H}+\hc\,,
\label{YukLagrangian_Fund}
\eea
which provide the following relations between Yukawa couplings and vevs:
\bea
Y_D \equiv \dfrac{\mean{\chi_d^L}\mean{\chi_d^{R\dag}}}{\Lambda_f^2}\,, \qquad \qquad 
Y_U \equiv \dfrac{\mean{\chi_u^L}\mean{\chi_u^{R\dag}}}{\Lambda_f^2}\,.
\label{YukawasVEVs_Bi_Fund_2F}
\eea
In this interesting case, the flavons are simply vectors under the flavour group, alike to quarks, with the 
simplest quantum number assignment being $\chi_{u,d}^L\sim(3,1,1)$, $\chi_u^R\sim(1,3,1)$ and 
$\chi_d^R\sim(1,1,3)$. Following this pattern, would the Yukawa couplings result from a condensate of fermionic 
flavons \cite{CG:MFV}, a $d=7$ Yukawa operator could be adequate
\beq
Y_D \equiv \dfrac{\mean{\ov{\Psi}_d^L\Psi_d^R}}{\Lambda_f^3}\,,\qquad \qquad 
Y_U \equiv \dfrac{\mean{\ov{\Psi}_u^L\Psi_u^R}}{\Lambda_f^3}\,,
\label{dim7Y}
\eeq 
with fermions quantum numbers under $G_f$ as in the previous case. 
Notice that these realizations in which the Yukawa couplings correspond to the vev of an aggregate of fields,  
rather than to a single field, are not the simplest realization of MFV as defined in Ref.~\cite{DGIS:MFV}, 
while still corresponding to the essential idea that the Yukawa spurions may have a dynamical origin. 

The goal of this work is to address the problem of the determination of the general scalar potential, compatible 
with the flavour symmetry $G_f$, for the flavon fields denoted above by $\Sigma$ or $\chi$. 
An interesting question is whether it is possible to obtain the SM Yukawa pattern - i.e. the observed values of quark 
masses and mixings- with a renormalizable potential. 
We derive the potential, analyze the possible vacua, and discuss the degree of ``naturalness'' of the possible solutions. 
It will be shown that the possibility of 
obtaining a large mass hierarchy and mixing at the renormalizable level varies much depending on the dimension of 
the Yukawa operator. The role played by non--renormalizable terms and the fine--tunings required to 
accommodate the full spectrum will be explored.

A relevant issue is what will be meant by natural: 
 following 't Hooft's naturalness 
 criteria, all dimensionless  free parameters of the potential not 
constrained by the symmetry should be of order one, and all dimensionful ones are expected to be of the order of the scale(s) of the 
theory. We will thus explore in which cases --if any-- those criteria allow that the minimum of the MFV potential 
corresponds automatically to mixings and large mass hierarchies. Stronger than \cal{O}($10$\%) adjustments (typical 
Clebsh-Gordan values in any theory) 
will be considered  fine-tuned.

It is worth to note that the structure of the scalar potentials constructed here is more general than the particular 
effective realization  in eqs.~(\ref{LagrangianBi_Fund}) and (\ref{YukLagrangian_Fund}). Indeed, it relies exclusively 
on invariance under the symmetry $G_f$ and on the flavon representation, bi-fundamental or fundamental{\footnote{For 
instance, the potential and the consequences for mixing obtained in this work will apply as well to the construction 
in ref.~\cite{GRV:SU3gauged}, notwithstanding the fact that there the  flavon vevs show and inverse hierarchy than that  for the minimal version of MFV, as they are proportional to the inverse of the SM Yukawa couplings.}

We limit our detailed discussions below to the quark sector. The implementation of MFV in the leptonic 
sector~\cite{CGIW:MLFV,DP:MLFV} requires some supplementary assumptions, as Majorana neutrino masses require 
to extend the SM and involve  a new scale: that of lepton number violation. Due to the smallness of neutrino 
masses, the {\it effective} scale of lepton number violation must be distinct from the flavour and electroweak 
ones, if new observable flavour effects are to be expected~\cite{GHHH:MFSM}. 
Nevertheless, the analysis of the flavon scalar potential performed below may also apply when considering 
leptons, although the precise analysis and implications for the leptonic spectrum will be carried 
out elsewhere.

The structure of the manuscript is as follows. In sect.~\ref{sec:2F}, for the two-family case we analyze the 
renormalizable potential for $d=5$ and $d=6$ Yukawa operators, or in other words of flavons in the bi-fundamental 
and in the fundamental of $G_f$, respectively, showing that in the latter case mixing and a strong hierarchy are 
intrinsically present. The corrections induced by non-renormalizable terms are also discussed. In sect.~\ref{sec:3F} 
the analogous analyses are carried out for the realistic three-family case and it is also discussed the qualitative new 
features appearing when considering simultaneously $d=5$ and $d=6$ Yukawa operators. The conclusions are 
presented in sect.~\ref{sec:concl}. Details of the analytical and numerical discussions of the potential minimization 
can be found in the Appendices.

%
%
\section{Two Family Case}
\label{sec:2F}

We start the discussion of the general scalar potential for the MFV framework by illustrating the two-family case,  
postponing the discussion of three families to the next section. Even if we restrict to a simplified case, with a smaller 
number of Yukawa couplings and mixing angles, it is a very reasonable starting-up scenario, that corresponds to the 
limit in which the third family is decoupled, as suggested by the hierarchy between quark masses and the smallness 
of the CKM mixing angles\footnote{We follow in this paper the PDG~\cite{PDG2010} conventions for the CKM matrix 
parametrization.} $\theta_{23}$ and $\theta_{13}$. In this section, moreover, we will introduce most of the conventions 
and ideas to be used later on for the three-family analysis.

With only two generations the non-Abelian flavour symmetry group, $G_f$, is reduced to
\beq
G_f=SU(2)_{Q_L}\times SU(2)_{U_R}\times SU(2)_{D_R}\,,
\eeq
under which the quark fields transform  as
\beq
Q_L\sim(2,1,1)\,,\qquad\qquad
U_R\sim(1,2,1)\,,\qquad\qquad
D_R\sim(1,1,2)\;.
\eeq
Following the MFV prescription, in order to preserve the flavour symmetry in the Lagrangian, the Yukawa spurions 
introduced in eq.~(\ref{spurions}) now transform under $G_f$ as
\beq
Y_U\sim(2,\ov{2},1)\,,\qquad\qquad 
Y_D\sim(2,1,\ov{2})\,.
\eeq
The masses of the first two generations and the mixing angle among them arise once the spurions take the following values:
\bea
Y_D=\left(
        \begin{array}{cc}
           y_d  & 0  \\
            0 & y_s  \\
        \end{array}
\right)\;,\qquad & & \qquad
Y_U={\mathcal V}_C^\dag\left(
        \begin{array}{cc}
           y_u  & 0  \\
            0 & y_c  \\
        \end{array}\right)\,, \label{SpurionsVEVs2F} 
\eea
where
\bea
{\mathcal V}_C =\left(
        \begin{array}{cc}
           \cos\theta & \sin\theta  \\
            -\sin\theta & \cos\theta  \\
        \end{array}
\right)
\label{Cabibbo}
\eea
is the usual Cabibbo rotation among the first two families.

%
%
\mathversion{bold}
\subsection{$d=5$ Yukawa operators: the bi-fundamental approach}
\label{sec:Bi_Fund2F}
\mathversion{normal}

The most intuitive approach, in looking for a dynamical origin of MFV, is probably to promote each Yukawa coupling
from a simple spurion to a flavon field. In other words, to consider the effective $d=5$ Lagrangian described 
above in eq.~(\ref{LagrangianBi_Fund}). The new fields --flavons-- are singlets under the SM gauge group but have, for the 
two-family case, the non-trivial transformation properties under $G_f$ given by
\beq
\Sigma_u\sim(2,\ov{2},1)\longrightarrow\Sigma_u'=\Omega_L\, \Sigma_u\, \Omega_{U_R}^\dag\;,\qquad
\Sigma_d\sim(2,1,\ov{2})\longrightarrow\Sigma_d'=\Omega_L\, \Sigma_d\, \Omega_{D_R}^\dag\,,
\label{FlavonsTransformations2F_Bifund}
\eeq
where $\Omega_X$ denotes the doublet transformation under the $SU(2)_X$-component of the flavour group. Once these 
flavon fields develop vevs as in eq.~(\ref{dim5Y}) and eq.~(\ref{SpurionsVEVs2F}), the flavour symmetry is explicitly 
broken and quark masses and mixings are originated. The effective field theory obtained at the electroweak scale is 
exactly MFV \cite{DGIS:MFV} (restricted to the two-family case). Then, within this approach, the problem of the origin 
of flavour is replaced by the need to explain if and how this particular vev configuration can naturally arise from 
the minimization of the associated scalar potential. 

This minimal framework can be easily extended in different ways, such as, for instance:
\begin{itemize} 

\item Considering different scales for the $\Sigma_u$ and $\Sigma_d$ flavon vevs.
\item Adding new representations. The most straightforward way to complete the basis in 
eq.~(\ref{FlavonsTransformations2F_Bifund}), is to add a third flavon transforming as a bi-fundamental 
of the RH components:
\beq
\Sigma_R \sim (1,2,\ov2) \longrightarrow \Sigma_R'=\Omega_{U_R}\, \Sigma_R\, \Omega_{D_R}^\dag\,.
\label{SigmaR}
\eeq
This new field does not contribute to the Yukawa terms, at least at the renormalizable level, but introduces new
operators with respect to MFV, which induce flavour changing neutral currents (FCNC)  mediating fully right-handed (RH) 
processes\footnote{The phenomenological impact of  these operators has already been introduced and studied in the 
three-family case in ref.~\cite{BGI:MFVwithRHcurrents}, in a different context.}.
\item Adding new replicas of the bi-fundamental representations.  
This could be very helpful as a natural source of new scales and possible mixings.
\end{itemize}
The first two possibilities do not affect essentially the flavour structure of the quark Yukawa couplings, which 
is the focus of this work, and we will not consider them below. No further consideration is given either in this section 
to the third possibility, both  for the sake of simplicity and because of the aesthetically unappealing aspect of being 
a trivial replacement of the puzzle of quark replication with that of flavon replication. 

We will thus restrict the remaining of this section to the analysis of the potential for just one $\Sigma_u$ and one 
$\Sigma_d$ fields, eqs.~(\ref{FlavonsTransformations2F_Bifund}). The general scalar potential, can then be written as a 
sum of two parts, the first dealing only with the SM Higgs fields and the second accounting also for the flavons interactions:
\beq
V\equiv V_H+V_\Sigma=-\mu^2 H^\dag H+\lambda_H (H^\dag H)^2+\sum_{i=4}^\infty V^{(i)}[H,\Sigma_u,\Sigma_d]\,.
\label{ScalarPotential}
\eeq
Inside $V^{(i)}$ all possible scalar potential terms of the effective field theory are included. In particular, 
$V^{(4)}$ contains all the renormalizable couplings written in terms of $H$ and $\Sigma_{u,d}$ while $V^{(i>4)}$  
incorporate all the non-renormalizable higher dimensional operators. There is no particular reason to impose that the 
EW and the flavour symmetry breaking should occur at the same scale. Indeed it is plausible that the flavour symmetry 
is broken by some new physics mechanism at a larger energy scale. Although it is true that the mixed Higgs-flavons terms 
could affect the value and location of the electroweak and flavour minima, the flavour composition of each term will not 
be modified by them. Once the flavour symmetry breaking occurs, all the terms in $V^{(i)}$ either contribute to the 
scalar potential as constants or can be redefined into $\mu^2$ or $\lambda_H$.  In what follows the analysis is 
restricted  to consider only the flavon part of the scalar potential, $V^{(i)}[\Sigma_u,\Sigma_d]$.

%
%
\subsubsection{The Scalar Potential at the Renormalizable Level}

From the transformation properties in eq.~(\ref{FlavonsTransformations2F_Bifund}), it is straightforward to write 
the most general independent invariants that enter in the scalar potential. At the renormalizable level, and for 
the case of two generations, five independent invariants can be constructed\footnote{Any other invariant operator 
can be expressed in terms of these five independent invariants. For example: $Tr\left(\Sigma_u\Sigma_u^\dagger
\Sigma_u\Sigma_u^\dagger\right)=Tr\left(\Sigma_u\Sigma_u^\dagger\right)^2-2\det\left(\Sigma_u\right)^2$.} 
\cite{Feldmann:2009dc}: 
\beq
\ba{ll}
A_u=\tr\left(\Sigma_u\Sigma_u^\dagger\right)\,,&\qquad B_u=\det\left(\Sigma_u\right)\,,\\[2mm]
A_d=\tr\left(\Sigma_d\Sigma_d^\dagger\right)\,,& \qquad B_d=\det\left(\Sigma_d\right)\,,\\[2mm]
A_{ud}=\tr\left(\Sigma_u\Sigma_u^\dagger\Sigma_d\Sigma_d^\dagger\right)\,.&
\ea
\label{Invariants2F_BiFund}
\eeq
Eqs. (\ref{dim5Y}) and (\ref{SpurionsVEVs2F})  allow to express these invariants in terms of physical 
observables, i.e. the four Yukawa eigenvalues and the Cabibbo angle: 
\bea
\mean{\Sigma_d} = \Lambda_f\,\left(
        \begin{array}{cc}
           y_d  & 0  \\
            0 & y_s  \\
        \end{array}
\right)\;,\qquad & & \qquad
\mean{\Sigma_u}=\Lambda_f\,{\mathcal V}_C^\dag\left(
        \begin{array}{cc}
           y_u  & 0  \\
            0 & y_c  \\
        \end{array}\right)\,, 
\label{SigmaSpurionsVEVs2F}
\eea
leading to:
\beq
\ba{l}
\mean{A_u}=\La^2\, (y_u^2+y_c^2)\,,\qquad \qquad \qquad \mean{B_u}=\La^2\, y_u\,y_c\,,\\[2mm]
\mean{A_d}=\La^2\, (y_d^2+y_s^2)\,,\qquad \qquad \qquad \mean{B_d}=\La^2\, y_d\,y_s\,,\\[2mm]
\mean{A_{ud}}=\La^4 \left[\left(y_c^2-y_u^2\right)\left(y_s^2-y_d^2\right)\cos2\theta+
        \left(y_c^2+y_u^2\right)\left(y_s^2+y_d^2\right)\right]/2\,.
\ea
\label{InvariantsExplicit2F_BiFund}
\eeq
Notice that the mixing angle appears only in the vev of $A_{ud}$, which is the only operator that mixes 
the up and down flavon sectors. This is as intuitively expected: {\it the mixing angle describes the relative
misalignment between two different directions in flavour space}. It is also interesting to notice that the expression 
for $\mean{A_{ud}}$ is related to the Jarlskog invariant for two families,  
\beq
4J=4\det{\left[Y_UY_U^\dagger,Y_DY_D^\dagger\right]}=\left(\sin{2\theta}\right)^2
       \left(y_c^2-y_u^2\right)^2\left(y_s^2-y_d^2\right)^2 \,,\nn
\eeq
by the following relation:
\beq
\dfrac{1}{\La^4}\dfrac{\derp \mean{A_{ud}}}{\derp \theta}= -2  \sqrt{J}\,.
\eeq
Using the invariants in eqs.~(\ref{Invariants2F_BiFund}), the most general renormalizable scalar 
potential allowed by the flavour symmetry reads:
\beq
V^{(4)}=\sum_{i=u,d}\left(-\mu_i^2A_i-\tilde{\mu}_i^2B_i+\lambda_iA_i^2+ 
        \tilde{\lambda}_i B_i^2\right)+g_{ud}A_uA_d+f_{ud}B_uB_d+\sum_{i,j=u,d}h_{ij}A_iB_j+\lambda_{ud}A_{ud}\,,
\label{ScalarPot2F_BiFund}
\eeq
where strict naturalness criteria would require all dimensionless couplings $\lambda$, $f$, $g$, $h$  to be of 
order $1$,  and the dimensionful $\mu$-terms to be 
smaller or equal than $\Lambda_f$ although of the same order of magnitude. It is clear from the start that, with the only 
use of symmetry implemented here,  a strict implementation of such criteria could lead at best to a strong hierarchy with 
some fields massless and the rest with masses of about the same scale. The ``fan" structure of quark mass splittings 
observed clearly calls, instead, for  a readjustment of the relative size of some  $\mu$ parameters, at least when 
restraining to the analysis of the renormalizable and classic terms of the potential. One question is whether, in this 
situation, even further fine-tunings are required among the mass parameters in the potential to accommodate nature.

The relations in eq.~(\ref{InvariantsExplicit2F_BiFund}) allow 
to determine the positions of the potential minima in terms of physical observables. A careful analytical and 
numerical study of the potential can be found in the Appendices. Here we briefly comment on the most  relevant 
physical results. Consider first the angular part of the potential. Deriving $V^{(4)}$ with 
respect to the angle $\theta$, it follows that
\beq
\left.\dfrac{\derp V^{(4)}}{\derp \theta}\right|_{min}\equiv \la_{ud}\dfrac{\derp{\mean{A_{ud}}}}{\derp \theta}\propto 
    \la_{ud}\sin{2\theta}\left(y_c^2-y_u^2\right)\left(y_s^2-y_d^2\right) \propto \lambda_{ud} \sqrt{J}\,.
\label{CabibboEq2F_BiFund}
\eeq
The minimum of the scalar potential thus occurs when at least one of the following conditions is satisfied 
i) $\la_{ud}=0$, ii) $\sin\theta=0$, iii) $\cos\theta=0$ or iv) two Yukawas in the same sector are degenerate. 
When condition i) is imposed, the angle remains undetermined;  
this assumption corresponds however to a severe fine-tuning on the model, as no symmetry protects this term 
from reappearing at the quantum level. Instead, due to the smallness of the Cabibbo angle, condition ii) can be 
interpreted as a first order solution which needs to be subsequently corrected, for example by the introduction 
of higher order operators. This possibility will be discussed in more detail in the next subsection. Finally, the last 
conditions, iii) and iv), are phenomenologically non representative of nature and large (higher order) corrections 
should be advocated in order to diminish the angle or to split the Yukawa degeneracy, respectively, making these 
solutions unattractive. All in all, the straightforward lesson that follows from eq.~(\ref{CabibboEq2F_BiFund}) 
is that, given the mass splittings observed in nature, {\it the scalar potential for bi-fundamental flavons does 
not allow  mixing at leading order.}

From the requirement that the derivatives of the scalar potential with respect to $y_{u,d,c,s}$ also vanish at the 
minima, four additional independent relations on the physical parameters are obtained. As discussed above, to obtain simultaneously a sizeable mixing and a mass spectrum largely splitted in masses, instead of generically degenerate, it is necessary to (re-)introduce a large, and unnatural, hierarchy among the 
different operators appearing in the scalar potential (see Appendix B for numerical details). 

These observations can be summarized stating that, with a natural choice of the coefficients appearing in  the 
renormalizable scalar potential $V^{(4)}$, after minimization one naturally ends up with a vanishing or undetermined 
mixing angle and with a naturally degenerate spectrum. 
In this respect we agree with a remark that can be found in refs.~\cite{Feldmann:2009dc,GRV:SU3gauged}. It is, however, 
interesting to notice that if the invariants $B_{u,d}$ (i.e. the determinant of the flavons) are neglected, which could 
be justified for example introducing some {\it ad hoc} discrete symmetry, the minima equations would then allow, instead, 
solutions non-degenerate in mass for same-charge quarks, with non-vanishing Yukawa couplings for the first (second) 
quark generations. This may open the possibility to study a modified version of the scalar potential in eq.~(\ref{ScalarPot2F_BiFund}), that predicts a natural 
hierarchy among the Yukawas of different generations.

%
%
\subsubsection{The Scalar Potential at the Non-Renormalizable Level}

Consider the addition of non-renormalizable operators to the scalar potential, $V^{(i>4)}$. It is very interesting to 
notice that this does {\it not} require the introduction of  new invariants beyond those in eq.~(\ref{Invariants2F_BiFund}): 
all higher order traces and determinants can in fact be expressed in terms of that basis of five ``renormalizable'' 
invariants. 

The lowest higher dimensional contributions to the scalar potential have dimension six (the complete list can be 
found in Appendix A). At this order, the only terms affecting the mixing angle are
\beq
V^{(6)}\supset \dfrac{1}{\Lambda_f^2}\sum_{i=u,d}\left(\alpha_iA_{ud}B_i+\beta_iA_{ud}A_i\right)\,.
\eeq
These terms, however, show the same dependence on the Cabibbo angle previously found  in eq.~(\ref{CabibboEq2F_BiFund}) 
and, consequently, they can simply be absorbed in the redefinition of the lowest order parameter, $\la_{ud}$. 
In other words, even at the non-renormalizable level, the most favorable trend leads to no mixing. To find a 
non-trivial angular structure it turns out that terms in the potential of dimension eight (or higher) have to be 
considered, that is
\beq
V^{(8)}\supset \lambda_{udud}A_{ud}^2\,,
\eeq
and eq.~(\ref{CabibboEq2F_BiFund}) would be replaced by
\beq
\left.\dfrac{\derp V}{\derp \theta}\right|_{min}\propto \sin{2\theta}\left(y_c^2-y_u^2\right)\left(y_s^2-y_d^2\right)
\left[\la_{ud}-2\,y^2_c y^2_s \la_{udud} \sin^2\theta + \dots \right]\,,
\eeq
implying
\beq
\sin^2\theta\simeq \dfrac{\la_{ud}}{2\, y^2_c y^2_s \la_{udud}}\,.
\eeq
Using the experimental values of the Yukawa couplings $y_s$ and $y_c$, a meaningful value for $\sin\theta$ can be 
obtained although at the price of assuming a highly fine-tuned hierarchy between the dimensionless coefficients of $d=4$ and 
$d=8$ terms, $\la_{ud}/\la_{udud}\sim 10^{-10}$, that cannot be naturally justified in an effective Lagrangian approach. 

The remaining four equations defining the minima, obtained deriving the scalar potential with respect to $y_{u,d,c,s}$, 
lead to no improvement  as compared to the renormalizable case: the Yukawa couplings are always given by general 
combinations of the coefficients of the scalar potential, underlining the complete absence of hierarchies among them. 
Realistic masses can be obtained at the classical level only when suitable fine-tunings are enforced~\footnote{See note added in proof}.

To summarize, it is possible to account for a non-vanishing mixing angle adding non-renormalizable terms to the scalar 
potential, although at the prize of introducing a large fine-tuning. This requirement comes in addition to the fact that 
the hierarchies among the Yukawa couplings can only be imposed by hand. Therefore the use of bi-fundamental scalar 
fields leads to an unsatisfactory answer to the problem of explaining the origin of flavour within the MFV hypothesis.

For the sake of illustrating the argument with a practical exercise, we conclude this section showing, as an explicit 
example, a fine-tuned scalar potential which can allow hierarchical Yukawas and a non-vanishing mixing angle:
\beq
V=\sum_i\left(-\mu_i^2A_i+\tilde{\lambda}_iB_i^2+\lambda_iA_i^2\right)+ 
\dfrac{\la_{udud}}{\Lambda_f^4}\left(A_{udud}-2A_{uudd}\right)- \epsilon_b \tilde{\mu}_d^2 B_d- 
\epsilon_u \tilde{\mu}_u^2 B_u+ \epsilon_\theta\lambda_{ud}A_{ud}\,,
\label{finetunepot}
\eeq
where $\epsilon_{u,d,\theta}$ are suppressing factors, 
which could justified via some discrete symmetry, and $A_{uudd}$, $A_{udud}$, dimension eight invariants defined by the following relations:
\beq
A_{udud}=\tr\left(\Sigma_u \Sigma^\dag_u \Sigma_d \Sigma^\dag_d \Sigma_u \Sigma^\dag_u \Sigma_d \Sigma_d^\dag\right)\,,\qquad
A_{uudd}=\tr\left(\Sigma_u \Sigma^\dag_u \Sigma_u \Sigma^\dag_u \Sigma_d \Sigma^\dag_d \Sigma_d \Sigma_d^\dag\right)\,.
\label{Invariants2F_BiFund1}
\eeq
By minimizing the potential in eq.~(\ref{finetunepot}) one obtains the following values for the Yukawa eigenvalues and 
the Cabibbo angle: 
\beq
\begin{gathered}
y_u \simeq \epsilon_u\,\dfrac{\sqrt{\lambda_u}\,\tilde\mu_u}{\sqrt2\,\tilde\lambda_u\,\mu_u}
\dfrac{\tilde\mu_u}{\Lambda_f}\,,\qquad\qquad
y_d \simeq \epsilon_d\, \dfrac{\sqrt{\lambda_d}\,\tilde\mu_d}{\sqrt2\,\tilde\lambda_d\,\mu_d}
\dfrac{\tilde\mu_d}{\Lambda_f}\,,\\[4mm]
y_c \simeq \dfrac{\mu_u}{\sqrt2\,\Lambda_f\,\sqrt\la_u}\,,\qquad\qquad
y_s \simeq \dfrac{\mu_d}{\sqrt2\,\Lambda_f\,\sqrt\la_d}\,,\\[4mm]
\sin^2\theta \simeq \epsilon_\theta\,\dfrac{\la_{ud}}{\la_{udud}\,y_c^2\,y_s^2}\,.
\end{gathered}
\eeq
Imposing for no good reason the values $\epsilon_u \sim 10^{-3}$, $\epsilon_d\sim5\times10^{-2}$, $\epsilon_\theta
\sim10^{-10}$ and $\mu/(\sqrt{\lambda}\Lambda_f)\approx\tilde\mu/(\sqrt{\tilde\lambda}\Lambda_f)\sim 10^{-3}$, the 
correct hierarchies between the quark masses and the correct Cabibbo angle could be obtained (see details for this 
special case in Appendix B). 

The discussion about $d=8$ terms presented above has pure illustrative purposes, as it may be a priori misleading to discuss 
the effects of $d=8$ terms in the potential without simultaneously considering quantum or other higher-order sources 
of corrections, such as the possible impact of a $\Sigma_R$ flavon\footnote{The impact of the fully RH bi-fundamental 
$\Sigma_R$ is  negligible: indeed it can enter in the scalar potential only as powers of $\Sigma_R\Sigma_R^\dag$ or 
its hermitian conjugate, and in particular, being a singlet of $SU(2)_{Q_L}$, it cannot mix with the other flavons. 
As a result, its contributions can always be absorbed through a redefinition of the parameters and then the conclusions 
above still hold.} - see eq.~(\ref{SigmaR}) -} or other $G_f$ representations.

%
%
\mathversion{bold}
\subsection{$d=6$ Yukawa operator: the fundamental approach}
\label{sec:Fund2F}
\mathversion{normal}

The identification of the Yukawa spurions as single flavon fields, transforming in the bi-fundamental representation 
of the flavour group (e.g. for a $d=5$ 
Yukawa operator), is only one of the possible ways the MFV ansatz can be implemented. An attractive alternative is to consider the Yukawas as composite objects or aggregates of several 
fields, e.g. suggesting Yukawa operators with $d>5$. In the simplest case, each Yukawa corresponds to two scalar fields $\chi$ transforming in the 
fundamental representation of $G_f$ (e.g. $Y\sim \mean{\chi} \mean{\chi'^\dagger}/\Lambda_f^2$, see 
eqs.~(\ref{YukLagrangian_Fund}) and (\ref{YukawasVEVs_Bi_Fund_2F})). This approach would {\it a priori} allow to  introduce  one 
new field for each component of the flavour symmetry: i.e. to reconstruct the spurions in eq.~(\ref{spurions}) just out 
of three vectors transforming as $(2,1,1)$, $(1,2,1)$ and $(1,1,2)$. However, such a minimal setup leads to an 
unsatisfactory realization of the flavour sector as no physical mixing angle is allowed at the renormalizable level~\footnote{ 
Because then the  flavons associated to the up and down left-handed character are not misaligned in flavour space, but correspond instead to 
just one $(2,1,1)$ flavon.}. The situation improves qualitatively, though, if two $(2,1,1)$ representations are introduced, 
one for the up and one for the down quark sectors. Consider then the following four fields:
\beq
\chi_u^L  \in  \left( 2,  1, 1\right)\,,\qquad\quad
\chi_u^R  \in  \left( 1, 2, 1\right)\,,\qquad\quad
\chi_d^L  \in  \left( 2,  1, 1\right)\,,\qquad\quad
\chi_d^R  \in  \left( 1,  1, 2\right)\,.
\label{listfundamentals}
\eeq
The corresponding $d=6$ effective Lagrangian  and Yukawa couplings have been shown in eqs.~(\ref{YukLagrangian_Fund}) 
and (\ref{YukawasVEVs_Bi_Fund_2F}). These flavons are vectors under the flavour symmetry. The only physical 
invariants that can be associated to vectors are the norm of the vectors and, eventually, their relative angles. 
Any matrix resulting from multiplying two vectors has only one non-vanishing eigenvalue, independently of the number 
of dimensions of the space. This fact alone already implies that, at the leading renormalizable order under discussion, 
just one ``up"-type quark and one ``down"-type quark are massive: a strong mass hierarchy between quarks of the same 
electric charge is thus automatic in this setup, which is a very promising first step in the path to explain the 
observed quark mass hierarchies. 

More in detail, the resulting Yukawa matrices are general $2\times2$ matrices, containing many unphysical parameters. Without loss 
of generality, it is possible to express the Yukawa couplings in terms of physical quantities by choosing the flavon 
vevs as follows: 
\beq
{\mean{\chi_i}}\equiv\left|\chi_i \right| {\mathcal V}_i
\left( 
\begin{array}{c} 
0 \\ 1
\end{array} 
\right)\,,
\label{VEV2F_Fund}
\eeq
 where  by $|\chi_i |$ we denote the norm of the vev of $\chi$, $|\chi_i | \equiv|\mean{\chi_i }|$, and ${\mathcal V}_i$ 
are $2\times2$ unitary matrices. 
 Redefining the quark fields as follows, 
\beq
Q_L'={\mathcal V}_L^{(d)\dag} Q_L\,,\qquad
U_R'={\mathcal V}_R^{(u)\dag} U_R\,,\qquad
D_R'={\mathcal V}_R^{(d)\dag} D_R\,,
\eeq
it results
\beq
\LL_Y=\overline{Q}_L'  Y_D D_R' H+ \overline{Q}_L'  Y_U U_R' \tilde{H} +\hc\,,
\eeq
with the corresponding Yukawa matrices given by~\footnote{The 
cutoff scale $\Lambda_f$ refers to the scale of the flavour dynamics. In principle we could have different scales 
for the left and right flavons as well as for the up and down ones, but for simplicity we assume that all the 
scales are  close and $\Lambda_f$ refers to the average value.}
\beq
Y_D=\dfrac {\left|\chi_d^L\right|\left|\chi_d^R\right|}{\Lambda_f^2} \left(\begin{array}{cc}
0 & 0\\
0 & 1
\end{array}\right)\,,\qquad\quad
Y_U=\dfrac{\left|\chi_u^L\right|\left|\chi_u^R\right|}{\Lambda_f^2}\, {\mathcal V}_L^{(d)\dag} {\mathcal V}_L^{(u)} 
\left(\begin{array}{cc}
0&0\\
0&1\end{array}\right)\,.
\label{Yukawas2F_Fund}
\eeq
This illustrates explicitly that: i) {\it there is a natural hierarchy among the mass of the first 
and second generations, without imposing any constraint on the parameters of the scalar potential}; ii) {\it the product 
${\mathcal V}_L^{(d)\dag} {\mathcal V}_L^{(u)}$ is a non-trivial unitary matrix that contains all the information about 
the mixing angle} (the phase can be easily removed in the two-family case under discussion). 
 There is now a clear geometrical 
interpretation of the Cabibbo angle: {\it the mixing angle between two generations of quarks is the misalignment 
of the $\chi^L$ flavons in the flavour space}, with the mixing matrix appearing in weak currents,  eq.~(\ref{Cabibbo}), given by
\beq
{\mathcal V_C}={\mathcal V}_L^{(u)\dag}\, {\mathcal V}_L^{(d)}\,.
\eeq 

Let us compare the phenomenology expected from bi-fundamental flavons (i.e. $d=5$ Yukawa operator) with that from fundamental 
flavons (i.e. $d=6$ Yukawa operators). For bi-fundamentals,  the  list of effective FCNC operators is exactly the same that 
in the original MFV proposal~\cite{DGIS:MFV}. The case of fundamentals presents some differences: higher-dimension invariants 
can be constructed in this case, exhibiting lower dimension than in the bi-fundamental case. 
 For instance, one can compare these two operators: 
\beq
\ov{D}_R\,\Sigma_d^\dag\,\Sigma_u\,\Sigma_u^\dag\,Q_L\sim[\text{mass}]^6\qquad \longleftrightarrow \qquad\ov{D}_R\,
\chi_d^{R}\,\chi_u^{L\dag}\,Q_L\sim[\text{mass}]^5\,,
\label{OperatorsFCNCdimensions}
\eeq
where the mass dimension of the invariant is shown in brackets; with  these two types of basic bilinear FCNC 
structures it is possible to build effective operators describing FCNC processes, but differing on the degree of 
suppression that they exhibit. 
This underlines the fact that the identification of Yukawa couplings with aggregates of two or more flavons is a 
setup which goes technically beyond the realization of MFV, resulting possibly in a distinct phenomenology which 
could provide a way to distinguish between fundamental and bi-fundamental origin.

%
%
\subsubsection{The scalar potential}
\label{sec:Fund2F_ScalarPot}

The general scalar potential that can be written including flavons in the fundamental is analogous to that in eq.~(\ref{ScalarPotential}),  replacing $\Sigma_i$ with $\chi_i$,
\beq
V\equiv V_H+V_\chi\,.
\eeq
Previous considerations regarding the scale separation between EW and flavour breaking scale hold also in this case, 
and in consequence the Higgs sector contributions will not be explicitly described. 

Any flavour invariant operator can be constructed out of the following five independent 
building blocks:
\beq
\chi_u^{L\dag}\chi_u^L\,,\qquad\qquad
\chi_u^{R\dag}\chi_u^R\,,\qquad\qquad
\chi_d^{L\dag}\chi_d^L\,,\qquad\qquad
\chi_d^{R\dag}\chi_d^R\,,\qquad\qquad
\chi_u^{L\dag}\chi_d^L\,.
\eeq
From the expressions for the Yukawa matrices in eqs.~(\ref{Yukawas2F_Fund}), it follows that in this scenario the scalar 
potential depends only on three of the five physical parameters: one angle and the two (larger) Yukawa couplings
\beq
\left|\chi_u^L\right|\left|\chi_u^R\right|=\Lambda_f^2 \, y_c\,,\qquad \qquad \,
\left|\chi_d^L\right|\left|\chi_d^R\right|=\Lambda_f^2 \, y_s\,,\qquad \qquad \,
\chi_u^{L\dagger}\chi_d^L=\cos\theta_c\left|\chi_u^L\right|\left|\chi_d^L\right|\,,
\label{connectionfund}
\eeq
given by the product of the left and right up (down) flavon moduli. As expected, the mixing angle is simply 
the angle defined in flavour space by the up and down left vectors. From the point of view of the measurable quantities, 
there is a certain parametrization freedom, and a possible convenient choice is given by\footnote{See Appendix C for 
a detailed discussion.}
\beq
\frac{\left|\chi_u^R\right|}{\Lambda_f}=\,1\,=\frac{\left|\chi_d^R\right|}{\Lambda_f}\,.
\label{RHvevs}
\eeq
As a result, the  invariants physically relevant for the flavour structure are:
\beq
\left|\chi_u^L\right|=\Lambda_f \,y_c\,,\qquad\quad 
\left|\chi_d^L\right|=\Lambda_f \, y_s\,,\qquad\quad 
\chi_u^{L\dag}\chi_d^L=\Lambda_f^2 y_c\,y_s\,\cos{\theta}\,.
\label{vevsFundamentals2D}
\eeq

At the renormalizable level, the scalar potential is  given by
\beq
V^{(4)}=-\sum_{i=u,d}\mu_i^2\, \chi_i^{L\dag}\chi_i^L-
\sum_{i=u,d}\tilde{\mu}_i^2\,\chi_i^{R\dag}\chi_i^R -
\mu_{ud}^2\,\chi_u^{L\dag}\chi_d^L+\ldots\,,
\eeq
where dots stand for all  possible quartic couplings. The total number of operators that can be introduced at 
the renormalizable level is 20. However, as shown in Appendix C, many of them (i.e. quartic couplings that mix 
different flavours) do not have any real impact on the existence and determination of the minima. 
Studying the latter, the following relations 
between the (large) up and down Yukawa eigenvalues and the Cabibbo angle follow:
\begin{equation}
\frac{y_{s}^2}{y_{c}^2}=\frac{\mu_d^2\lambda_u}{\mu_u^2\lambda_d}\, , \qquad \qquad \cos\theta=
\frac{\sqrt{\lambda_u\lambda_d}\mu_{ud}^2}{\lambda_{ud}\mu_u\mu_d}\,
\label{YukawasResults}
\end{equation}
which shows that without strong fine-tunings this scenario can explain the hierarchy between the first and second 
family, and account for a sizable Cabibbo angle.

%
%
\subsubsection{The first generation}
\label{sec:Fund2F_FirstGen}

In this two-generation analysis, the first family has remained massless at the renormalizable level. A first 
possibility is that non-renormalizable corrections may induce this small masses.
Non-renormalizable interactions manifest themselves in form of higher order contributions to the Yukawa operators 
and the flavon vevs and/or as non-renormalizable terms in the potential, which can modify its minima. 

From eq.~(\ref{YukLagrangian_Fund}) and the flavon transformation properties, it follows that higher order 
contributions to the Yukawa operators can only be constructed by further insertions of $\chi^\dag\chi$ inside the 
renormalizable operators. However, such kind of insertions do not modify the flavour structure of the Yukawa 
matrices, but simply redefine the two heavier couplings, $y_c$ and $y_s$.
On the other hand, the introduction of higher order operators in the scalar potential has the effect of modifying  
the vevs of the flavons, replacing the relation in eq.~(\ref{VEV2F_Fund}) with 
\beq
\dfrac{\mean{\chi_{u,d}^{L,R}}}{\Lambda_f}\equiv\left|(1+\cO(\epsilon))\chi_{u,d}^{L,R}\right|
\left(\mathcal V_{L,R}^{(u,d)}(1+\cO(\epsilon) )
\right)
\left(\begin{array}{c}
\cO(\epsilon) \\ 1
\end{array}\right)\,,
\eeq
where $\epsilon\ll1$ parametrizes the ratio among higher and leading order contributions. The only effect of these 
modifications is to redefine the mixing angle $\theta$ and the second family Yukawas, $y_c$ and $y_s$, without 
changing the rank of the Yukawa matrices and leaving thus the first generation massless. In summary, non-renormalizable 
interactions cannot switch on additional (first family) Yukawas if they were absent at the renormalizable level.

An alternative can be built on the fact that each up-down set of fundamental flavons provides a supplementary scale, 
in addition to new sources of mixing from their misalignment. A possibility along this direction is to enlarge 
the number of flavons to six, made out of a set of three ($\chi_{u,d}^R$ plus just one $\chi^L$)
replicated:  in total 
two $\chi^L\sim(2,1,1)$, two $\chi_u^R\sim(1,2,1)$ and two $\chi_d^R\sim(1,1,2)$. 
In this case the Yukawa terms change in a non-trivial way:
\beq
Y_D\equiv\dfrac{\sum_{ij}\alpha^d_{ij}\mean{\chi_i^L}\mean{\chi_j^{R\dag}}}{\Lambda_f^2}\,, \qquad \qquad 
Y_U\equiv\dfrac{\sum_{ij}\alpha^u_{ij}\mean{\chi_i^L}\mean{\chi_j^{R\dag}}}{\Lambda_f^2}\,,
\eeq
with $\alpha_{i,j}$ numerical coefficients and $i,j$ running over all flavons. An explicit computation reveals that, 
for generic values of $\alpha_{ij} (\neq 0)$, the rank of the Yukawa matrices is indeed two. However, in this case, 
the natural hierarchy between the first and second family is lost, being all the Yukawas of the same order unless 
the vevs of the new flavons are unnaturally smaller than those of the first replica.  In conclusion, adding new RH 
flavon copies does not lead either to an appealing  and natural source of masses for the first generation.

%
%
\section{The Three-Family Case}
\label{sec:3F}

Let us extend the previous analysis to the three-family case. 
While most of the procedure, with both bi-fundamental and fundamental representations, follows straightforwardly, two main 
differences should be underlined. First of all, the top Yukawa coupling, $y_t$, is now a parameter which is of $\cO(1)$. 
The fact that in the two-family case the largest Yukawa, $y_c$ was much smaller than one, allowed us to safely 
retain only the lowest order terms in the (Yukawa) perturbative expansion. In the three-family scenario, in 
principle, one should include all orders in the expansion. However, in this case, the Cayley-Hamilton identity~\cite{CNS:SUSYMFVRunning,MS:SUSYMFVEDM} provides a way out, as it proves that 
a general $3\times3$ matrix $X$ must satisfy the relation:
\beq
X^3-\tr[X]\, X^2+\dfrac{1}{2}X\left(\tr[X]^2-\tr\left[X^2\right]\right)-\det[X]=0\,,
\eeq 
which allows to express all powers $X^n$ (with $n>2$) in terms solely of $\unity$, $X$ and $X^2$,  
with coefficients involving the traces of $X$ and $X^2$ and the determinant of $X$. In the case under study, 
$X$ corresponds to the invariant products $\Sigma^\dag \Sigma$ or $\chi^\dag \chi$, depending on whether bi-fundamental 
or fundamental representations are considered. 

The second main difference with respect to the two-family case, is the appearance of a physical phase in the 
quark mixing matrix. For the sake of simplicity, in this paper we disregard CP-violation,
 deferring its discussion  to a future work\cite{AGMR:CPpaper}.

%
%
\mathversion{bold}
\subsection{$d=5$ Yukawa operator: the bi-fundamental approach}
\label{sec:Bi_Fund3F}
\mathversion{normal}
In this section we extend the approach discussed in sect.~\ref{sec:Bi_Fund2F} to the three-family case. 
 Consider two bi-triplets under the flavour symmetry $G_f$, see eq.~(\ref{spurions}),
\beq
\Sigma_u\sim(3,\ov{3},1)\longrightarrow\Sigma_u'=\Omega_L\, \Sigma_u\, \Omega_{U_R}^\dag\;,\qquad
\Sigma_d\sim(3,1,\ov{3})\longrightarrow\Sigma_d'=\Omega_L\, \Sigma_d\, \Omega_{D_R}^\dag\,,
\label{FlavonsTransformations3F_Bifund}
\eeq
where now the $\Omega_X$ matrices refer to the triplet transformations under the $SU(3)_X$ component of the flavour 
group. The Yukawa Lagrangian is the same as that in eq.~(\ref{LagrangianBi_Fund}). Once the flavons develop a vev 
as in eq.~(\ref{dim5Y}), the flavour symmetry is broken and one recovers the observed fermion masses and CKM matrix 
given in eq.~(\ref{SpurionsVEVs}).  
Recall that the present realization is the simplest realization of the original MFV approach \cite{DGIS:MFV}. 
Again, it would be possible to extend it introducing a third RH flavon field, $\Sigma_R\sim (1,3,\ov3)\longrightarrow\Sigma_R'=\Omega_{U_R}\, \Sigma_R\, 
\Omega_{D_R}^\dag$. We do not further consider it when constructing the scalar potential, as it cannot contribute  
to the Yukawa spurions neither at $\cO(1/\Lambda_f)$ nor $\cO(1/\Lambda_f^2)$, that is, neither via $d=5$ nor $d=6$ 
Yukawa operators. 

Restricting the explicit  analysis to the part of the renormalizable scalar potential not containing the 
SM Higgs field, a complete and independent basis is given by the following seven invariant operators:
\beq
\ba{ll}
A_u=\tr\left(\Sigma_u\Sigma_u^\dagger\right)\,,& \qquad\qquad 
\mean{A_u}=\Lambda_f^2\left(y_t^2+y_c^2+y_u^2\right)\,,\\[2mm]
B_u=\det\left(\Sigma_u\right)\,,& \qquad \qquad
\mean{B_u}=\Lambda_f^3\,y_u\,y_c\,y_t\,,\\[2mm]
A_d=\tr\left(\Sigma_d\Sigma_d^\dagger\right)\,,& \qquad \qquad
\mean{A_d}=\Lambda_f^2\left(y_b^2+y_s^2+y_d^2\right)\,,\\[2mm]
B_d=\det\left(\Sigma_d\right)\,,& \qquad \qquad
\mean{B_d}=\Lambda_f^3\,y_d\,y_s\,y_b\,,\\[2mm]
A_{uu}=\tr\left(\Sigma_u\Sigma_u^\dagger\Sigma_u\Sigma_u^\dagger\right)\,,& \qquad \qquad
\mean{A_{uu}}=\Lambda_f^4\left(y_t^4+y_c^4+y_u^4\right)\,,\\[2mm]
A_{dd}=\tr\left(\Sigma_d\Sigma_d^\dagger\Sigma_d\Sigma_d^\dagger\right)\,,& \qquad \qquad
\mean{A_{dd}}=\Lambda_f^4\left(y_b^4+y_s^4+y_d^4\right)\,,\\[2mm]
A_{ud}=\tr\left(\Sigma_u\Sigma_u^\dagger\Sigma_d\Sigma_d^\dagger\right)\,,& \qquad \qquad
\mean{A_{ud}}=\Lambda_f^4\left(P_0+P_{int}\right)\,,
\ea
\label{Invariant3F_Fund}
\eeq
where $P_0$ and $P_{int}$ encode the angular dependence,
\bea
&&P_0\equiv-\sum_{i<j} \left(y_{u_i}^2-y_{u_j}^2 \right)  \left( y_{d_i}^2-y_{d_j}^2 \right) \sin^2 \theta_{ij}\,,
\label{P0}\\
&&\begin{split}
P_{int}\equiv&
\sum_{i<j,k}\left(y_{d_i}^2-y_{d_k}^2\right)\left(y_{u_j}^2-y_{u_k}^2\right)\sin^2 \theta_{ik}\sin^2\theta_{jk} \, +\\
&-\left( y_d^2-y_s^2 \right)\left( y_c^2-y_t^2 \right) \sin^2\theta_{12} \sin^2\theta_{13} \sin^2\theta_{23} \,+ \\[3mm]
&+\dfrac{1}{2}\left(y_d^2-y_s^2\right)\left(y_c^2-y_t^2\right)\cos\delta\,\sin2\theta_{12}\sin2\theta_{23}\sin\theta_{13}\,,
\end{split}
\label{Pint}
\eea
with $i,j,k=1,2,3$. $P_0$ generalizes the expression found in the two-family case - see eq.~(\ref{InvariantsExplicit2F_BiFund}) - containing all the terms 
with a single angular dependence. The second piece, instead, $P_{int}$, contains all contributions that involve more 
than one mixing angle. Notice that in this case the Jarlskog invariant appears only at the non-renormalizable level.

The most general scalar potential at the renormalizable level is now given by
\beq
V^{(4)}=\sum_{i=u,d}\left(-\mu_i^2A_i+\tilde{\mu}_iB_i+\lambda_iA_i^2+\lambda'_iA_{ii} \right)+g_{ud}A_uA_d + 
\lambda_{ud}A_{ud}\,.
\label{newV4Bifund}
\eeq
Notice that the invariants $B_{u,d}$ have mass dimension three (instead of two for the two-generation case), so that no $B^2_{u,d}$ term can be introduced  at this level.

The solutions that minimize this scalar potential have a pattern very similar 
to that for the two-family case: i) no mixing is favored~\footnote{However, due to the peculiar structure of the last term in eq.~(\ref{Pint}), minima with 
non-vanishing angles are now allowed, although leading to solutions which are both fine-tuned and overall physically incorrect.} , ii) all eigenvalues 
tend to be degenerate in most of the parameter space. Now however, there is 
a region in parameter space for which a hierachical solution is allowed for non 
strictly zero, but constrained, $\tilde \mu$ . This solution has one nonvanishng Yukawa 
eigenvalue per up and down sector, but to recover the hierarchy among top 
and bottom masses it is necessary to  further demand $g_{ud} < y^2_b /y_t^2$ which, in the absence of {\it ad hoc} symmetries,  results in a 
similar  degree of �ne-tunnig to that for the two-family case. Furthermore, alike
to the case of an initial  vanishing $\sin \theta$ at the renormalizable level for two 
families, it cannot be corrected by non-renormalizeble 
terms in the potential. 

As in sect. \ref{sec:Bi_Fund2F} for two generation, we studied the contributions of non-renormalizable operators in the 
scalar potential, with similar conclusion: the introduction of higher order terms does not lead to a more natural 
description of the physical parameters. Nevertheless, some improvement can be obtained when discussing the scenario 
with a fine-tuned  choice of the parameters $g_{ud}$ and $\tilde{\mu}$. In this case, in fact, lighter Yukawas can 
be introduced through higher order operators, even if no natural hierarchy between the first two families can be obtained.

In summary, for three generations, to consider bi-fundamental scalars (as in the case of $d=5$ Yukawa operator) alone as 
the possible dynamical origin of Yukawa couplings  does not lead naturally to a satisfactory pattern of masses and mixings~\footnote{See note added in proof}.

%
%
\mathversion{bold}
\subsection{$d=6$ Yukawa operator: the fundamental approach}
\label{sec:Fund3F}
\mathversion{normal}

We deal now with the case of flavons transforming in the fundamental of the flavour group $G_f$. For most of the 
conventions we refer to the two-family treatment done in sect.~\ref{sec:Fund2F}.
To account for non-trivial mixing, it is necessary to introduce at least four flavons, corresponding to up and down, 
left and right flavons:
\beq
\chi_u^L  \in  \left( 3,  1, 1\right)\,,\qquad\quad
\chi_u^R  \in  \left(1, 3, 1\right)\,,\qquad\quad
\chi_d^L  \in  \left( 3,  1, 1\right)\,,\qquad\quad
\chi_d^R  \in  \left(1,  1, 3\right)\,.
\label{listfundamentals3d}
\eeq
When they develop vevs, the flavour symmetry is spontaneously broken and the Yukawa matrices 
are given as in eq.~(\ref{YukawasVEVs_Bi_Fund_2F}). 
Without loss of generality, it is possible to write:
\beq
\mean{\chi_{u,d}^{L,R}} \equiv\left|\chi_{u,d}^{L,R}\right|\mathcal{V}_{L,R}^{(u,d)}
\left(\begin{array}{c}
		0\\
		0\\
		1
\end{array}\right)\,,
\label{VEV3F_Fund}
\eeq
where $\mathcal{V}_{L,R}^{(u,d)}$ are $3\times3$ unitary matrices. Similarly to what was shown in sect.~\ref{sec:Fund2F}, 
removing the unphysical parameters, the following expressions for the Yukawa matrices are obtained:
\beq
Y_D=\frac{\left|\chi_d^L\right|\left|\chi_d^R\right|}{\Lambda_f^2} \left(\begin{array}{ccc}
0 & 0 & 0\\
0 & 0 & 0\\
0 & 0 & 1\\ 
\end{array}\right)\,,\qquad\quad																	
Y_U=\frac{\left|\chi_u^L\right|\left|\chi_u^R\right|} {\Lambda_f^2} \mathcal{V}_L^{(d)\dag} \mathcal{V}_L^{(u)} 
\left(\begin{array}{ccc}
0 & 0 & 0\\
0 & 0 & 0\\
0 & 0 & 1\\ 
\end{array}\right)\,.
\label{Yukawas3F_Fund}
\eeq
This illustrates that, independently of the parametrization chosen,  $Y_D$ and $Y_U$ can have only one non-vanishing 
eigenvalue, as they result from multiplying two vectors. For obvious reasons, in eq.~(\ref{Yukawas3F_Fund}) the massive 
state is chosen to be that of the third generation.
  The flavon vevs have not  broken completely the flavour symmetry, leaving a residual $SU(2)_{Q_L}\times SU(2)_{D_R}\times 
SU(2)_{U_R}$ symmetry group. As a consequence any rotation in the $12$ sector is unphysical and the only physical 
angle, given by the misalignment between $\mean{\chi_u^L}$ and $\mean{\chi_d^L}$ in the flavour space, can be identified 
with the $23$ CKM mixing angle: 
\beq
\mathcal{V}_L^{(d)\dag} \mathcal{V}_L^{(u)}=\left(
\begin{array}{ccc}
1&0&0\\
0&\cos{\theta_{23}}&\sin{\theta_{23}}\\
0&-\sin{\theta_{23}}&\cos{\theta_{23}}\\
\end{array}\right)\,.
\eeq 
The analysis of the scalar 
potential follows exactly that in sect.~\ref{sec:Fund2F} for two families (see for example eq.~(\ref{YukawasResults})),
 with the obvious replacement of $y_c$, $y_s$ for $y_t$, $y_b$ and with the physical mixing 
angle corresponding now to $\theta_{23}$. Both the largest hierarchy and a $\cos \theta_{23}$ naturally of $\cO(1)$
are beautifully explained without any fine-tuning. However, as in the two-family case, it is not possible to generate 
lighter fermion masses either introducing non-renormalizable interactions or adding extra RH flavons.

Nevertheless, the partial breaking of flavour symmetry provided by eq.~({\ref{Yukawas3F_Fund}}) can open 
quite interesting possibilities from a model-building point of view. Consider as an example the following 
multi-step approach. In a first step, only the minimal number of fundamental fields are introduced: i.e. $\chi^L$, $\chi^R_u$ 
and $\chi^R_d$. Their vevs break $G_f=SU(3)^3$ down to $SU(2)^3$, originating non-vanishing Yukawa couplings 
only for the top and the bottom quarks, without any mixing angle (as we have only one left-handed flavon). As a second step, 
 four new $G_f$-triplet fields $\chi^{\prime L,R}_{u,d}$ are added,  whose contributions to the Yukawa terms are suppressed 
relatively to the previous flavons (i.e. $\mean \chi' \ll \mean \chi$). If their vevs 
 point in the direction of the unbroken flavour subgroup 
$SU(2)^3$, then the residual symmetry is further reduced. As a result, non-vanishing charm and strange Yukawa couplings 
are generated together with a mixing among the first two generations:
\beq
\begin{aligned}
&Y_u\equiv \frac{\mean{\chi^L}\,\mean{\chi_u^{R\dag}}}{\Lambda_f^2}+ \frac{\mean{\chi_u^{\prime L}}\,
\mean{\chi_u^{\prime R\dag}}}{\Lambda_f^2}=\left(\begin{array}{ccc}
0&\sin{\theta}\,y_c&0\\
0&\cos{\theta}\,y_c&0\\
0&0&y_t\\
\end{array}\right)\,,\\
&Y_d\equiv \frac{\mean{\chi^L}\,\mean{\chi_d^{R\dag}}}{\Lambda_f^2}+ \frac{\mean{\chi_d^{\prime L}}\,
\mean{\chi_d^{\prime R\dag}}}{\Lambda_f^2}=\left(\begin{array}{ccc}
0&0&0\\
0&y_s&0\\
0&0&y_b\\
\end{array}\right)\,.
\end{aligned}
\eeq
The relative suppression of the two sets of flavon vevs correspond to the hierarchy between $y_c$ and $y_t$ ($y_s$ and $y_b$)\footnote{Alternatively, all flavon vevs of similar magnitude with different flavour scale would lead to the same pattern.}. Hopefully, a 
refinement of this argument would allow to explain the rest of the Yukawas and the remaining angles.  The 
construction of the scalar potential for such a setup would be quite model dependent though, and beyond the scope of this paper.

%
%
\subsection{Combining fundamentals and bi-fundamentals}
\label{sec:Bi_Fund-Fund}

Until now we have considered separately Yukawa operators of dimension $d=5$ and $d=6$. It is, however, interesting 
to explore if some added value from the simultaneous presence of both kinds of operators can be obtained. 
This is a sensible choice from the point of view of effective Lagrangians in which, working at $\cO(1/\Lambda_f^2)$, contributions of four types may be included: i) the leading $d=5$ $\cO(1/\Lambda_f)$ operators; ii) renormalizable terms stemming from fundamentals (i.e. from $d=6$ $\cO(1/\Lambda_f^2)$ operators); iii) $\cO(1/\Lambda_f^2)$ of the form $\Sigma_{u,d} \Sigma_R$ if $\Sigma_R$ turns out to be present in the spectrum; iv) other corrections numerically competitive at the orders considered here. We focus here as illustration on the impact of i) and ii):
\beq
\LL_Y=\ov{Q}_L\left[\dfrac{\Sigma_d}{\Lambda_f}+
\dfrac{\chi_d^L\chi_d^{R\dagger} }{\La^2} \right]D_RH +
\ov{Q}_L\left[\dfrac{\Sigma_u}{\Lambda_f}+
\dfrac{\chi_u^L \chi_u^{R\dagger}}{\La^2}  \right]U_R\tilde{H}+\hc\,,
\label{LagrangianMixed}
\eeq
As the bi-fundamental flavons arise at first order in the $1/\La$  expansion, it is suggestive to think of the fundamental 
contributions as a ``higher order'' correction.  
Let us then consider the case in which 
the flavons develop vevs as follows:
\beq
\dfrac{\mean{\Sigma_{u,d}}}{\Lambda_f}\sim\left(
        \begin{array}{ccc}
           0  & 0 & 0 \\
           0  & 0 & 0 \\
           0  & 0 & y_{t,b} \\
        \end{array}
\right)\,,\qquad\qquad
\dfrac{\mean{\chi_{u,d}^L}}{\Lambda_f^2}\sim \left(
        \begin{array}{ccc}
           0 \\
           y_{c,s} \\
           0 \\
        \end{array}
\right)\,,
\label{lastequation}
\eeq
and $\chi_{u,d}^R$ acquire arbitrary vev values, although $\cO(1)$, for all components.  Nevertheless, it is important 
to recall that the bi-fundamentals $\Sigma$ point in most cases to degenerate Yukawa eigenvalues instead of the pattern 
in the left-hand side of eq.~(\ref{lastequation}), and either restrictive conditions on the parameters, or an extra 
symmetry, have to be imposed to obtain it, see sects. 2.1.1 and 3.2. Finally, 
\begin{equation}
\begin{aligned}
&Y_u=\left(\begin{array}{ccc}
0&\sin{\theta_c}\,y_c&0\\
0&\cos{\theta_c}\,y_c&0\\
0&0&y_t\\
\end{array}\right)\,,\qquad
&Y_d
\left(\begin{array}{ccc}
0&0&0\\
0&y_s&0\\
0&0&y_b\\
\end{array}\right)\,.
\end{aligned}
\end{equation}
This seems an appealing pattern, with masses for the two heavier generations and one sizable mixing angle, that we chose to identify here with the Cabibbo angle\footnote{Similar constructions have been suggested also in other contexts as in \cite{Berezhiani:2005tp,FKR:AnarchicalMasses,CFRZ:UnifiedAnarchicalMasses}.}. As for the lighter family,  non-vanishing masses for the up and  down quarks could now result from non-renormalizable operators.

The drawback of these combined analysis is that the direct connection between the minima of the potential and the spectrum is lost and the analysis of the potential would be very involved.

%
%

\section{Conclusions}
\label{sec:concl}

The ansatz of MFV implicitly assumes a dynamical origin for the SM Yukawa couplings. In this paper we explored 
such a possibility. The simplest dynamical realization of MFV 
is to identify the Yukawa couplings with the vevs of some dynamical fields, the flavons. For instance, 
the Yukawa interactions themselves could result, after spontaneous symmetry breaking, from effective operators of 
dimension $d>4$ invariant under the flavour symmetry, which involve one or more flavons together with the usual SM fields. 
 
Only a scalar field (or an aggregate of fields in a scalar configuration) can get a vev, which should correspond 
to the minimum of a potential. What may be the scalar potential of the MFV flavons? May some of its minima naturally correspond to the SM spectra of masses and mixing angles? These are the questions addressed in this work. 

First of all, we showed here that the underlying flavour symmetry - under which the terms in the potential have to be 
invariant -  is a very restrictive constraint: at the renormalizable level only a few terms are allowed in the potential, 
and even at the non-renormalizable level quite constrained patterns have to be respected.

The simplest realization is obtained by a one-to-one correspondence of  each Yukawa coupling with a single scalar 
field transforming 
in the bi-fundamental of the flavour group. In the language of effective Lagrangians, this may correspond to the 
lowest order terms in the flavour expansion: $d=5$ effective Yukawa operators made out of one flavon field plus  
the usual SM fields. We have constructed the general scalar potential for bi-fundamental flavons, both for the case of 
two and three families. At the renormalizable level, at the minimum of the potential only vanishing or undetermined 
mixing angles are allowed. The introduction of either additional {\it ad hoc} symmetries or the restriction to a contrived region of the parameter domain could allow 
to obtain solutions with vanishing Yukawa couplings for all quarks but those in the heaviest family. Still, mixing would 
be absent. The addition of non-renormalizable terms to the potential would allow masses  for the lighter families, although 
without providing naturally a correct pattern of masses and mixings.
In resume,  the sole consideration of flavons in the bi-fundamental representation of the flavour group does not naturally 
lead to a satisfactory dynamical description of the SM quark flavour sector, at least at the classical level.

Another avenue explored in this work  associates two vector flavons to each Yukawa spurion, i.e. a Yukawa 
$Y\sim \mean{\chi^L}\mean{\chi^{R\dag}}/\Lambda_f^2$. This is a very attractive scenario in that while 
Yukawas are composite objects, the new fields are in the fundamental representation of the flavour group, 
in analogy with the  case of quarks. Those flavons could be scalars or fermions: we focused exclusively on scalars. 
From the point of view of effective Lagrangians, this case could correspond to the next-to leading order term 
in the expansion:  $d=6$ Yukawa operators. We have constructed the general scalar potential for scalar flavons 
in the fundamental representation, both for the case of two and three families of quarks. By construction, this scenario 
results unavoidably in a strong hierarchy of masses: at the renormalizable level only one quark gets mass in each quark 
sector: they could be associated with the top and bottom quark for instance. Non-trivial mixing requires as expected a 
misalignment between the flavons associated to the up and down left-handed quarks. In consequence, the minimal field 
content corresponds to four fields $\chi^L_u$, $\chi^L_d$, $\chi^R_u$ and $\chi^R_d$, and the physics of mixing lies 
in the interplay of the first two. In resume, for fundamental flavons it follows in a completely natural way: 
i) a strong mass hierarchy between quarks of the same charge, pointing to a distinctly heavier quark in each sector; 
ii) one non-vanishing mixing angle, which can be identified with the Cabibbo angle in the case of two generations, 
and for instance with the rotation in the $23$ sector of the CKM matrix in the case of three generations.

Nevertheless, to achieve non-vanishing Yukawa couplings for the lighter quarks and the full mixing pattern requires, 
at least at the classical level explored here, more complicated scenarios and variable degrees of fine-tuning. 
Interesting possibilities which we started to explore here include replicas of fundamental flavons, in several varieties. 
An intriguing one consists in considering the minimal set of $\chi$ fields plus their replicas: it allows a double step symmetry breaking 
mechanism, which may produce the hierarchical quark spectrum and the shell-like pattern of the CKM matrix.

Finally, we briefly explored the possibility of introducing simultaneously bi-fundamen\-tals and fundamentals flavons. 
It is a very sensible possibility from the point of view of effective Lagrangians to consider both  $d=5$ and 
$d=6$ Yukawa operators when working to $\cO(1/\Lambda_f^2)$. It suggests that $d=5$ operators, which bring in the 
bi-fundamentals, could give the dominant contributions, while the $d=6$ operator  - which brings in the fundamentals - 
should provide a correction inducing the masses of the two lighter families and the Cabibbo angle. It requires, though, 
to appeal to a discrete symmetry or to restrict the parameters to the potential to a contrived region to avoid quark 
mass degeneracies induced by the bi-fundamental flavons.
 
Overall, it is remarkable  that  the requirement of invariance under the flavour symmetry strongly constraints 
the scalar potential of MFV, up to the point that the obtention of quark mass hierarchies and mixing angles is far 
from trivial. Furthermore, besides exploring the - disappointing - impact in mixing of bi-fundamental flavons, 
this work has shown that flavons in the fundamental are instead a tantalizing avenue to induce hierarchies and 
non-trivial fermion mixing. A long path remains ahead, though, to naturally account for the complete observed 
fermion spectrum and mixings.

%
%

\section*{Acknowledgements}

We are specially indebted to Enrico Nardi for fruitful  discussions and suggestions. We also thank 
Alvaro de Rujula and Pilar Hern\'andez for illuminating discussions. L. Merlo and S. Rigolin thank 
the Departamento de F\'isica Te\'orica of the Universidad Aut\'onoma de Madrid for hospitality during 
the development of this project. 
R. Alonso and M.B. Gavela acknowledge CICYT through the project FPA2009-09017 and by CAM through the project 
HEPHACOS, P-ESP-00346. R. Alonso acknowledges financial support from the MICINN grant BES-2010-037869. 
L. Merlo acknowledges the German `Bundesministerium f\"ur Bildung und Forschung' under contract 05H09WOE. 
S. Rigolin acknowledges the partial support of an Excellence Grant of Fondazione Cariparo and of the 
European Program‚ Unification in the LHC era‚ under the contract PITN- GA-2009-237920 (UNILHC).

\section*{Note added in proof}
After this work was submitted, a paper appeared in the arXiv \cite{Nardi:2011st} where it has been suggested that 
the introduction of Coleman-Weinberg quantum corrections to our results for the bi-fundamental case could generate 
subdominant Yukawa splittings.

The author, using a slightly modified version of our notation, re-derived our renormalizable potential for the bi-fundamental case for three families. Looking at eq.~(\ref{newV4Bifund}) and eqs.~(\ref{Invariants2F_BiFund})--(\ref{ScalarPot2F_BiFund}), it is easy to identify the relations to move from one notation to the other:
\begin{eqnarray}
&& A_u \rightarrow T_u \, , \quad B_u \rightarrow D_u \, , \quad A'_{uu} = (A_u^2 - A_{uu}) \rightarrow 2\, A_u
\nonumber \\
&& \mu_u\rightarrow m_u \, , \quad \tilde{\mu}_u\rightarrow 2\,|\tilde{\mu}_u| \, , \quad \lambda_u\rightarrow 
\lambda_u+\dfrac{1}{2}\tilde{\lambda}'_u \, , \quad \lambda'_u\rightarrow -\dfrac{1}{2}\tilde{\lambda}'_u\,.
\nonumber
\end{eqnarray}
The freedom on the relative sign between the determinant and $\tilde{\mu}$ terms allowed in our paper has been retaken in v2 of \cite{Nardi:2011st} as a 
cosine dependence, which now allows negative coefficients and redefines their norm.

The hierarchical solution in which that paper is based, was already identified in our work, together with the 
degenerate one. To quantify the validity range of the two solutions we found, we add here a detailed analysis 
of the stability of the potential, that was not included in our previous version. 

\begin{description}
\item A) {\it Stability condition for two families.} \\
For the two-family case the extremality equations read: 
\begin{equation}
 \frac{\partial V_u}{\partial y_c}=4\,\lambda_u\,\Lambda_{fl}^2\,y_c\left(A_u-\frac{\mu^2_u}{2\,\lambda_u}\right)+
 2\,\tilde{\lambda}_u\,\Lambda_{fl}^2\,y_u\left(B_u-\frac{\tilde{\mu}_u^2}{2\,\tilde{\lambda}_u}\right)-h_u\,\Lambda_{fl}^2\left(
y_u\,A_u+2\,y_c\,B_u\right)=0 \nonumber
\end{equation}
\begin{equation}
 \frac{\partial V_u}{\partial y_u}=4\,\lambda_u\,\Lambda_{fl}^2\,y_u\left(A_u-\frac{\mu^2_u}{2\,\lambda_u}\right)+
 2\,\tilde{\lambda}_u\,\Lambda_{fl}^2\,y_c\left(B_u-\frac{\tilde{\mu}^2_u}{2\,\tilde{\lambda}_u}\right)-h_u\,\Lambda_{fl}^2
 \left(y_c\,A_u\,+\,2\,y_u\,B_u\right)=0 \, .\nonumber
\end{equation}
One can easily verify that the hierarchical pattern $(0,\,y)$ is not a solution of these equations, 
unless a severe fine-tuning on the parameters $\mu_u$, $\tilde \mu_u$, $\lambda_u$ and $h_u$ is introduced. 
Only the symmetric solution $(y,\,y)$ arises as a natural minimum. 

\item B) {\it Stability condition for three families.} \\
The conditions defining the minima now read as follows.
\begin{enumerate}
 \item  The parameter region in which only the symmetric solution $(y,\,y,\,y)$ provides a stable minimum is defined by
 \begin{equation}
 \frac{\tilde{\mu}^2_u}{\mu^2_u}>\frac{8\,\lambda_u'^2}{\lambda_u\,+\,\lambda'_u}\, 
 \nonumber
 \end{equation}
 for $\lambda'_u < 0$. On the other hand, for $\lambda'_u > 0$, the configuration $(y,\,y,\,y)$ is a stable minimum for any value of $\tilde{\mu}^2_u/\mu^2_u$.
 
 \item The parameter region in which the symmetric solution is the absolute minimum, while the hierarchical configuration $(0,\,0,\,y)$ is a local minimum corresponds to
 \begin{equation} 
 8\left(\lambda_u+\lambda_u'\right)\left(\left(4-2\frac{\lambda'_u}{\lambda_u+\lambda_u'}\right)^{3/2}
 -\left(8-6\frac{\lambda_u'}{\lambda_u+\lambda_u'}\right)\right) < 
\frac{\tilde{\mu}^2_u}{\mu^2_u}<\frac{8\,\lambda_u'^2}{\lambda_u+\lambda'_u}
 \end{equation}  
 
 \item The parameter region in which the symmetric solution is a local minimum, while the hierarchical solution is the absolute minimum is defined by
  \begin{equation} \frac{8\,\lambda_u'^2}{3\lambda_u+2\lambda'_u}<\frac{\tilde{\mu}^2_u}{\mu^2_u}<
 8\left(\lambda_u+\lambda_u'\right)\left(\left(4-2\frac{\lambda'_u}{\lambda_u+\lambda_u'}\right)^{3/2}
 -\left(8-6\frac{\lambda_u'}{\lambda_u+\lambda_u'}\right)\right)\,.
 \end{equation}
 
 \item Finally, the parameter region in which only the hierarchical configuration is a minimum corresponds to
 \begin{equation}
 \frac{\tilde{\mu}^2_u}{\mu^2_u}<\frac{8\,\lambda_u'^2}{3\lambda_u+2\lambda'_u}\,.
 \end{equation}
\end{enumerate}

\begin{figure}[t]
\centering
 \includegraphics[width=4in]{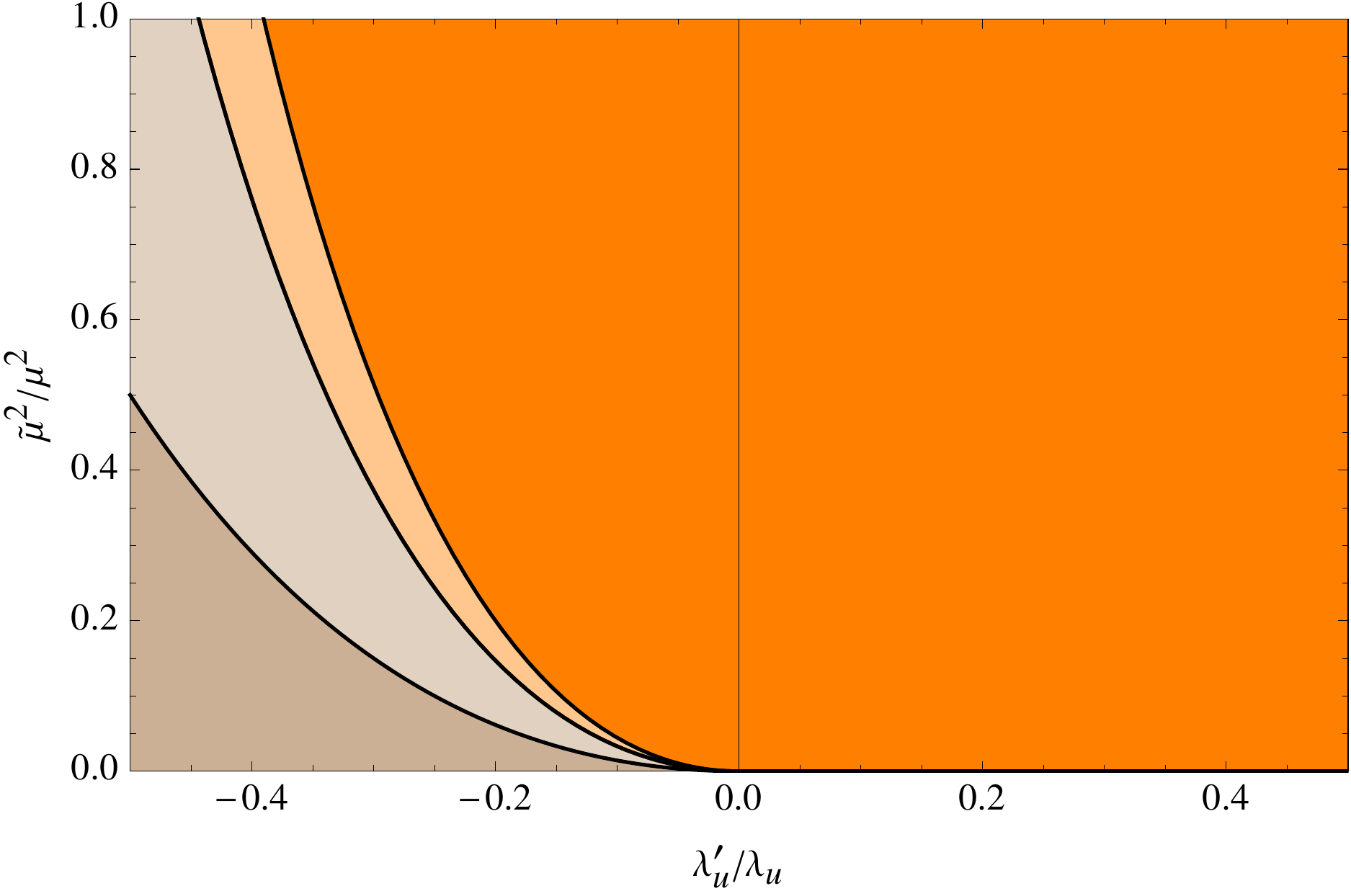}
 \caption{Parameter space for the symmetric and hierarchical configurations. The dark-Orange corresponds to a region where the symmetric 
  solution $(y,y,y)$ is the stable absolute minimum, while the hierarchical solution $(0,0,y)$ is a saddle point. In the light-Orange 
  region, the symmetric configuration is the absolute minimum, while the hierarchical solution is a local one. On the contrary, in the light-Brown 
  region the symmetric configuration is a local minimum, while the hierarchical solution is an absolute one. Finally, in the dark-Brown region only the hierarchical solution is a minimum. In the plot the value $\lambda_u=1/2$ has been used for illustration.}
\label{fig:note}
\end{figure}

\end{description}

As illustrated in fig.~(\ref{fig:note}) (see also \cite{Planck2011}), for a typical $\lambda_u$ value in the perturbative 
regime the symmetric configuration is the absolute minimum for most of the parameter space (here shown in dark and light orange). However, even when the hierarchical solution is preferred, yielding non-zero top and bottom Yukawa eigenvalues only, the hierarchy among up and down sectors must be considered. In particular the presence of the term $g_{ud}\,A_{u}\,A_{d}$ must be constrained by setting $g_{ud}\,<\,y_b^2\,/\,y_t^2\,\sim\,10^{-3}$  to warranty the top-bottom mass hierarchy. As already stated before, such a fine-tuning can be justified through additional symmetries. In particular in \cite{Nardi:2011st}, it is placed in the vev of flavons transforming under Abelian factors.

\appendix

 %
 %

\mathversion{bold}
\appendixA{A \hspace{0.2cm}  d=6 operators in the bifundamental approach}
\mathversion{normal}
 
We give here a summary of the invariant operators that appear in the potential at dimension 6 for the Bi-fundamental approach. 
There are two types of operators: products of the invariants in eq. (\ref{Invariants2F_BiFund}) and new operators written in terms of traces of flavons. The list of the former invariants is:
\begin{equation}
\begin{gathered}
\tr\left(\Sigma_i\Sigma_i^\dagger\right)\tr\left(\Sigma_j\Sigma_j^\dagger\right)\tr\left(\Sigma_k\Sigma_k^\dagger\right)\,,
\qquad\det\left(\Sigma_i\right)\tr\left(\Sigma_j\Sigma_j^\dagger\right)\tr\left(\Sigma_k\Sigma_k^\dagger\right)\,,\\
\det\left(\Sigma_i\right)\det\left(\Sigma_j\right)\tr\left(\Sigma_k\Sigma_k^\dagger\right)\,,\qquad
\det\left(\Sigma_i\right)\det\left(\Sigma_j\right)\det\left(\Sigma_k\right)\,,\\
\tr\left(\Sigma_i\Sigma_i^\dagger\right)\tr\left(\Sigma_u\Sigma_u^\dagger\Sigma_d\Sigma_d^\dagger\right)\,,\qquad
\det\left(\Sigma_i\right)\tr\left(\Sigma_u\Sigma_u^\dagger\Sigma_d\Sigma_d^\dagger\right)\,.
\end{gathered}
\end{equation}
where $i$, $j$ and $k$ run over u,d. The new invariant operators that appear are of the form
\begin{equation}
\tr\left(\Sigma_i\Sigma_i^\dagger\Sigma_j\Sigma_j^\dagger\Sigma_k\Sigma_k^\dagger\right)\,.
\end{equation}

In the two family case the vevs of these operators are not independent and can be expressed as linear combinations 
of the lowest order (LO) ones. To understand it, notice that five parameters (four Yukawas and an angle) suffice to parametrize 
the vevs. Then the relations of these parameters with the first five LO invariants can be formally inverted and substituted in any 
higher dimension new invariant, to express them as functions of the five former invariants. As an example, 
\[
\begin{split}
\tr\left(\mean{\Sigma_u}\mean{\Sigma_u^\dagger}\mean{\Sigma_u}\mean{\Sigma_u^\dagger}\mean{\Sigma_d}\mean{\Sigma_d^\dagger}\right)=&\tr\left(\mean{\Sigma_u}\mean{\Sigma_u^\dagger}\right)\tr\left(\mean{\Sigma_u}\mean{\Sigma_u^\dagger}\mean{\Sigma_d}\mean{\Sigma_d^\dagger}\right)+\\
&-\det\left(\mean{\Sigma_u}\right)^2\tr\left(\mean{\Sigma_d}\mean{\Sigma_d^\dagger}\right)\,.
\end{split}
\]

 %
 %

 \mathversion{bold}
 \appendixB{B \hspace{0.2cm} A fine-tuned scalar potential in the bifundamental approach}
 \mathversion{normal}

\begin{figure}[h]
	\centering
	\includegraphics[width=3in]{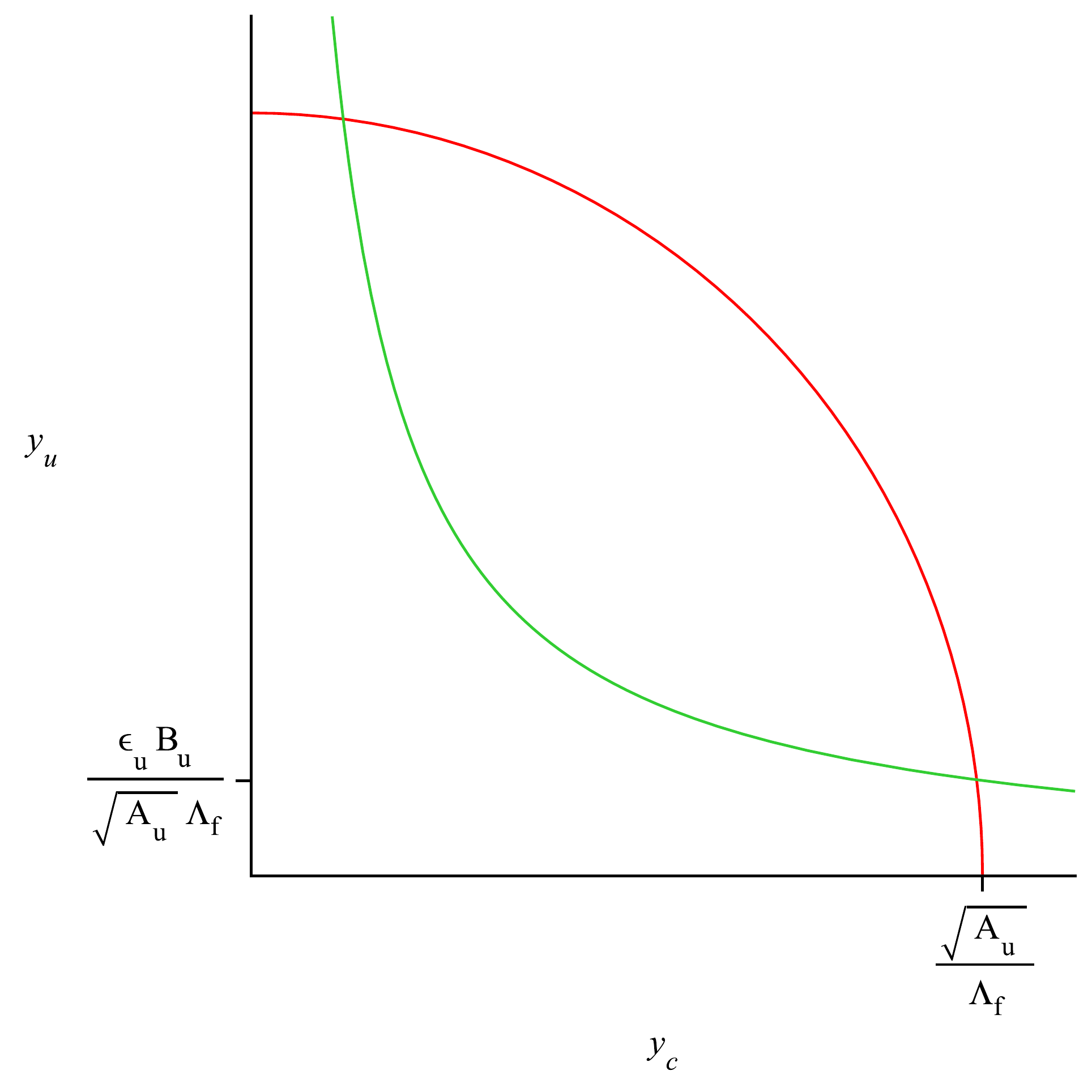}
	\caption{Graphic determination of the minimum.}
	\label{grapview}
\end{figure}

This Appendix gives details on a particular scalar potential whose minimum sets the observed values of masses and mixings for the first two generations. Its purpose is to illustrate the theoretical prize to be paid in order to obtain a realistic solution, and to discuss its degree of naturalness. This ansatz for the potential is given in eq. (\ref{finetunepot}). Rewriting it as a sum of four double well potential terms involving the Yukawa eigenvalues and terms involving the mixing angle, it reads:
\begin{equation}
V_\Sigma=\sum_{i=u,d}\left[\lambda_i\left(A_i-\frac{\mu_i^2}{2\lambda_i}\right)^2+\tilde{\lambda}_i\left( B_i-\epsilon_i\frac{\tilde{\mu}_i^2}{2\tilde{\lambda}_i}\right)^2\right]+\frac{\lambda_{udud}}{\Lambda_f^4}\left(A_{udud}-2A_{uudd}\right)+\epsilon_\theta\lambda_{ud}A_{ud}\,.
\label{proppot}
\end{equation}
Here $\epsilon_u$, $\epsilon_d$ and $\epsilon_\theta$ parametrize the suppression of the respective operators and will be defined when discussing the results. The invariants in the previous expression have already been defined in eqs. (\ref{Invariants2F_BiFund}), (\ref{Invariants2F_BiFund1}), and the renormalizable operators have been expressed in terms of masses and mixings in eq. (\ref{InvariantsExplicit2F_BiFund}). At the minimum of the potential, the non-renormalizable term corresponds to:
\begin{equation}
\begin{gathered}
\begin{split}
\mean{A_{udud}-2A_{uudd}}=&\Lambda_f^8\left[\left(y_c^2-y_u^2\right)^2\left(y_s^2-y_d^2\right)^2\sin^4\theta+\right.\\
&\hspace{8mm}\left.+2\left(y_c^2y_d^2+y_s^2y_u^2\right)\left(y_c^2-y_u^2\right)\left(y_s^2-y_d^2\right)\sin^2\theta-y_c^4y_s^4-y_d^4y_u^4\right]\,.
\end{split}
\end{gathered}
\label{defprop}
\end{equation}
The first part of the potential being positive definite, it is minimized when vanishing, which implies $\left\langle A_u\right\rangle=\Lambda_f^2(y_c^2+y_u^2)=\mu_u^2/2\lambda_u$,  $\left\langle B_u\right\rangle=\Lambda_fy_cy_u=\epsilon_u\tilde{\mu}_u^2/2\tilde{\lambda}_u$ and similar expressions for the down sector. These equations define a circle and an hyperbola in the $(y_c,y_u)$ plane. Their intersection defines the minimum as depicted in figure \ref{grapview}.

\subsection*{Minimization of the scalar potential}
 The explicit equations for the minimum of the scalar potential considered above are shown in what follows. In a first approximation we neglect all the terms suppressed by $\epsilon_{u,d,\theta}$.
\begin{itemize}
\item 
The equation associated with $y_u$ is then given by:
\begin{equation}
\dfrac{\partial V_\Sigma}{\partial y_u}=2y_u\Lambda_f^4\left[-\frac{\mu_u^2}{\Lambda_f^2}+2\lambda_u\left(y_u^2+y_c^2\right)+\tilde{\lambda}_u y_c^2-2\lambda_{udud}F_u\right]=0\,.
\label{eqminyu}
\end{equation}
The term $F_u$ is a function of parameters that will not enter in the determination of $y_u$ and vanishes in the limit of massless first family and no mixing. 
  The physical choice in eq. (\ref{eqminyu}) is to cancel the first factor taking $y_u=0$, as  the cancellation of the other factor would lead to $y_c=0$. This is stable provided that $\tilde{\lambda}_u>0$. In a similar way, $y_d=0$ is a solution to the equation $\partial V_ \Sigma/\partial y_d=0$ .

\item 
When deriving with respect to the angle $\theta$ we find
\begin{equation}
\begin{split}
\dfrac{\partial V_\Sigma}{\partial \theta}=&2\sin{2\theta}\,\lambda_{udud}\left(y_c^2-y_u^2\right)\left(y_s^2-y_d^2\right)\times\\
&\times\left[\left(y_c^2-y_u^2\right)\left(y_s^2-y_d^2\right)\sin^2\theta+\left(y_c^2y_d^2+y_s^2y_u^2\right)\right]=0\,.
\end{split}
\label{eqanglec}
\end{equation}
Substituting the solutions to the previous minima equations considered, $y_u=0=y_d$, eq. (\ref{eqanglec}) forces $\sin{2\theta}=0$.

\item 
For the heavy Yukawa couplings, once $y_u=y_d=0$ is chosen the equations take the form:
\begin{equation}
\frac{\partial V_\Sigma}{\partial y_c}=2y_c\,\Lambda_f^4\left(2\,\lambda_u\,y_c^2-2\,\lambda_{udud}\,y_c^2\,y_s^4-\frac{\mu_u^2}{\Lambda_f^2}\right)=0\,.
\end{equation}
Neglecting the trivial solution, which is unstable for positive definite coefficients, this  equation yields the expression for $y_c$:
\begin{equation}
y_c^2=\frac{\mu_u^2}{2\Lambda_f^2\left(\lambda_u-\lambda_{udud}\,y_s^4\right)}\simeq\frac{\mu_u^2}{2\Lambda_f^2\lambda_u}\,,
\end{equation}
where the last equality holds when taking into account the observed value of the strange Yukawa coupling. A similar result can be found for $y_s$.
\end{itemize}

Summarizing, neglecting all terms  suppressed by $\epsilon_{u,d,\theta}$, the minimum of the scalar potential is given by:
\begin{equation}
y_u=y_d=0\,,\qquad \quad
y_c=\dfrac{\mu_u}{\sqrt{2}\Lambda_f\sqrt{\lambda_u}}\,,\qquad \quad
y_s=\dfrac{\mu_d}{\sqrt{2}\Lambda_f\sqrt{\lambda_d}}\,,\qquad \quad
\sin\theta=0\,.
\end{equation}
The observed values of $y_c$ and $y_s$ Y are understood as the outcome of the hierarchy among the vevs of the flavons, $\langle\chi\rangle\sim\mu$, and the flavour scale $\Lambda_f$. Note that the parameters $\epsilon_{u,d,\theta}$ do not enter into the definition of $y_c$ and $y_s$, but control the hierarchy between the light and the heavy generations and the appearance of  a non-trivial mixing angle. This solution is stable with all the coefficients in eq. (\ref{proppot}) positive and furthermore the inclusion of the corrections given by the $\epsilon$-terms will shift the minimum but will not change its stability.\\

We now discuss the changes of the solutions found above by the introduction of $\epsilon_{u,d,\theta}$.

\begin{itemize}
\item 
The corrections for the first family Yukawa couplings shift their values from zero by an amount $\epsilon_{u,d}$. Explicitly, once the leading order solution found for $y_c$ and $y_s$ is inserted into eq. (\ref{eqminyu}), the dominant contributions are given by
\begin{equation}
\dfrac{\partial V_\Sigma}{\partial y_u}=\Lambda_f^4\left[2y_u\tilde{\lambda}_u y_c^2-\epsilon_u\frac{\tilde{\mu}_u^2}{\Lambda_f^2}y_c\right]=0\,,
\end{equation}
which leads to a non-vanishing Yukawa coupling for the up quark:
\begin{equation}
y_u=\epsilon_u\frac{\sqrt{\lambda_u}\,\tilde{\mu}_u}{\sqrt{2}\,\tilde{\lambda}_u\,\mu_u}\frac{\tilde\mu_u}{\Lambda_f}\,.
\end{equation}
A similar result holds also for $y_d$.

\item 
When considering the equation that determines the mixing angle, several corrections are present, although the dominant one is given by
\begin{equation}
\dfrac{\partial V_\Sigma}{\partial \theta}=2\sin{\theta}\cos{\theta}y_c^2y_s^2\left[2\lambda_{udud}\left(y_c^2y_s^2\sin^2\theta\right)-\epsilon_\theta\lambda_{ud}\right]=0\,,
\end{equation}
and the corresponding non-trivial solution reads
\begin{equation}
\sin^2\theta=\epsilon_\theta\dfrac{\lambda_{ud}}{2\,\lambda_{udud}\,y_s^2\,y_c^2}\,.
\end{equation}
\end{itemize}

The minimum of the scalar potential proposed in eq. (\ref{proppot}) is then given by
\begin{eqnarray}
\begin{gathered}
y_u \simeq \epsilon_u\frac{\sqrt{\lambda_u}\,\tilde{\mu}_u}{\sqrt{2}\,\tilde{\lambda}_u\,\mu_u}\frac{\tilde\mu_u}{\Lambda_f}\,,\qquad\qquad
y_d \simeq \epsilon_d\frac{\sqrt{\lambda_d}\,\tilde{\mu}_d}{\sqrt{2}\,\tilde{\lambda}_d\,\mu_d}\frac{\tilde\mu_d}{\Lambda_f}\,,\\[4mm]
y_c \simeq \frac{\mu_u}{\sqrt2\,\Lambda_f\,\sqrt\lambda_u}\,,\qquad\qquad
y_s \simeq \frac{\mu_d}{\sqrt2\,\Lambda_f\,\sqrt\lambda_d}\,,\\[4mm]
\sin^2\theta \simeq \epsilon_\theta\dfrac{\lambda_{ud}}{2\,\lambda_{udud}\,y_s^2\,y_c^2}\,.
\end{gathered}
\label{solbifund}
\end{eqnarray}

We can now specify the value of $\epsilon_{u,d,\theta}$ in order to accommodate the observed hierarchies and mixing for the first two generations: considering the ratios $\mu/(\sqrt\lambda\Lambda_f)\approx\tilde\mu/(\sqrt{\tilde\lambda}\Lambda_f)\sim10^{-3}$, 
it follows that
\begin{equation}
\epsilon_u\sim 10^{-3}\,,\qquad \qquad
\epsilon_d\sim 5\times10^{-2}\,,\qquad \qquad
\epsilon_\theta\sim 10^{-10}\,,
\end{equation}
must hold.
A comment is in order: when discussing this special illustrative scalar potential, we considered up to dimension 8 operators, while neglecting many terms otherwise allowed by the symmetry. However, even such an arbitrary choice was not sufficient to recover realistic mass hierarchies and the mixing angle, and  further fine-tunings were required, including $\epsilon$ values as tiny as $10^{-10}$  to recover the Cabibbo angle. These remarks should suffice to show how unnatural is the set up when trying to fix all observables from pure $d=5$ Yukawa operators.

\subsection*{Three Family Case}
The three family case involves a wider variety of operators. This is because some of the accidental simplifications in two families no longer hold for three. The analytic treatment to find the minima becomes more complicated as well, as the number of observables increases to six quark masses and three angles (obviating the CP-odd phase). We present a graphic analysis of the scalar potential in this case. This approach assumes a positive definite potential whose minimum is just the point in which the geometrical surfaces defined by constant invariant quantities meet. When focusing on the masses in either the up or the down sector, we project the parameter space to one that has as many dimensions as families. This means that instead of the curves in the $(y_c,y_u)$ plane of figure \ref{grapview} we will consider surfaces in $(y_t,y_c,y_u)$ space.

The lowest dimension invariants that involve Yukawa eigenvalues only for the up sector correspond to: 
\begin{equation}
\begin{array}{l}
\mean {A_u}=\Lambda_f^2\,\left(y_t^2+y_c^2+y_u^2\right)\,,\\[2mm]
\mean {B_u}=\Lambda_f^3\,y_t\,y_c\,y_u\,,\\[2mm]
\mean {A'_{uu}}=\mean{A_u^2-A_{uu}}=2\,\Lambda_f^4\left(y_t^2\,y_c^2+y_u^2\,y_t^2+y_c^2\,y_u^2\right)\,,
\end{array}
\label{inv3fam}
\end{equation}
where the last invariant is introduced as a linear combination of some of those in eq. (\ref{Invariant3F_Fund}). Notice that three independent invariants are necessary to fix the three different masses. We can study the intersection of the surfaces defined by giving fixed values to these operators.

\begin{figure}[h]
\begin{minipage}{0.3 \textwidth} 
\begin{center}
	\includegraphics[width=1.5in]{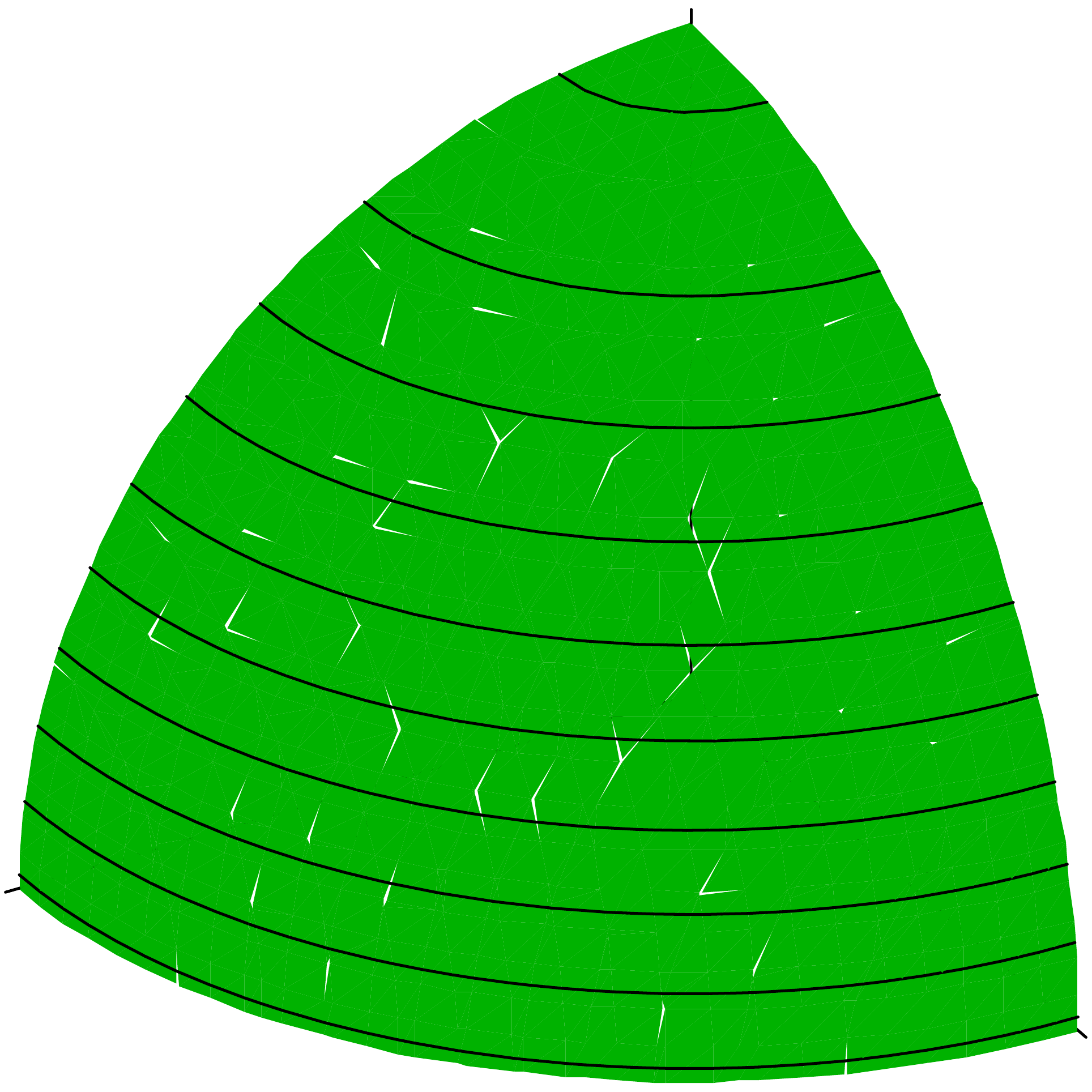}
	\caption{Surface of constant $A_{u}$ in $\{y_u, y_c, y_t\}$ space.}
	\label{fig1}
\end{center}
\end{minipage}
\hfill \begin{minipage}{0.3\textwidth}
\begin{center}
	\includegraphics[width=1.5in]{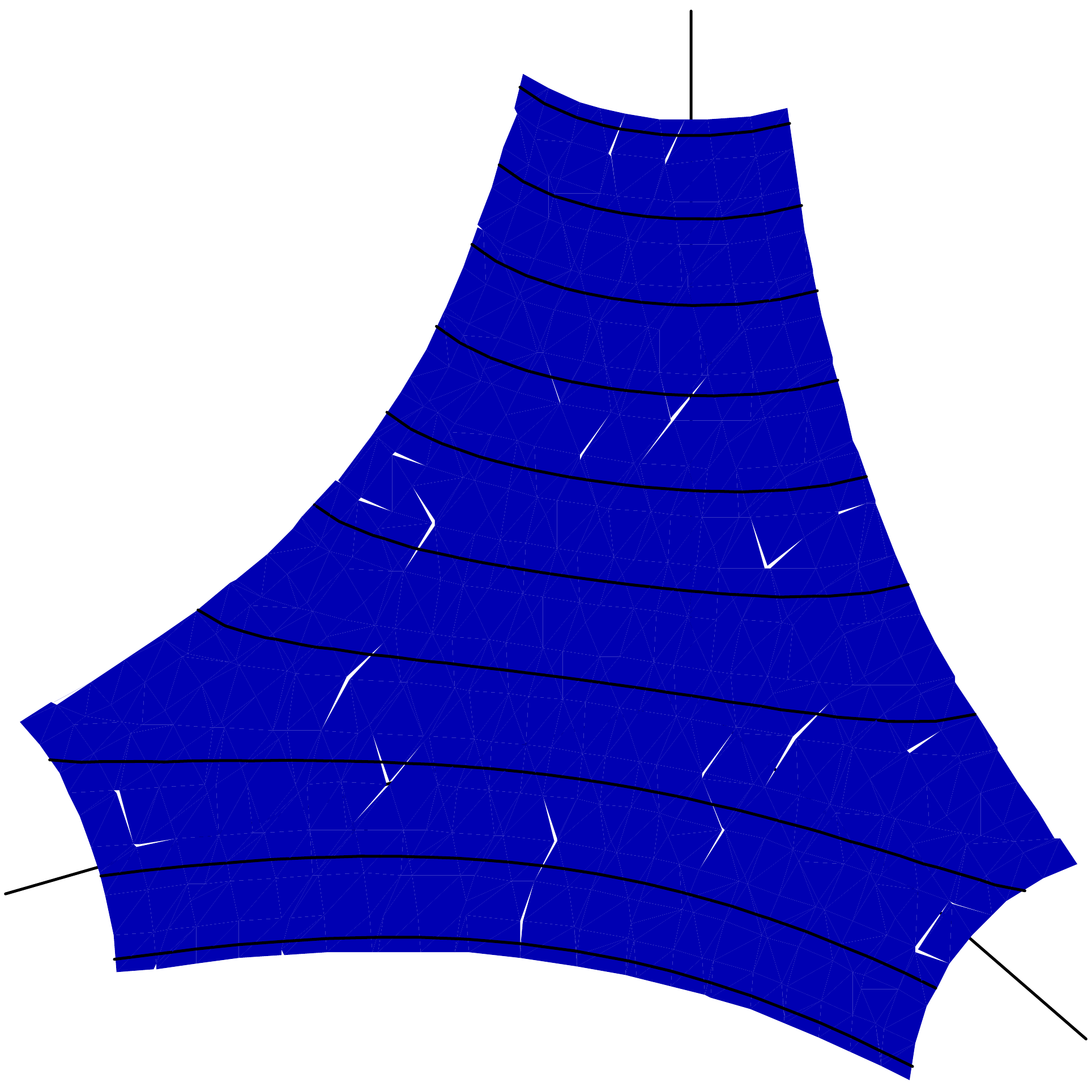}
	\caption{Surface of constant $A'_{uu}$ in $\{y_u, y_c, y_t\}$ space.}
	\label{fig2}
\end{center}
\end{minipage}
\hfill \begin{minipage}{0.3\textwidth}
\begin{center}
	\includegraphics[width=1.5in]{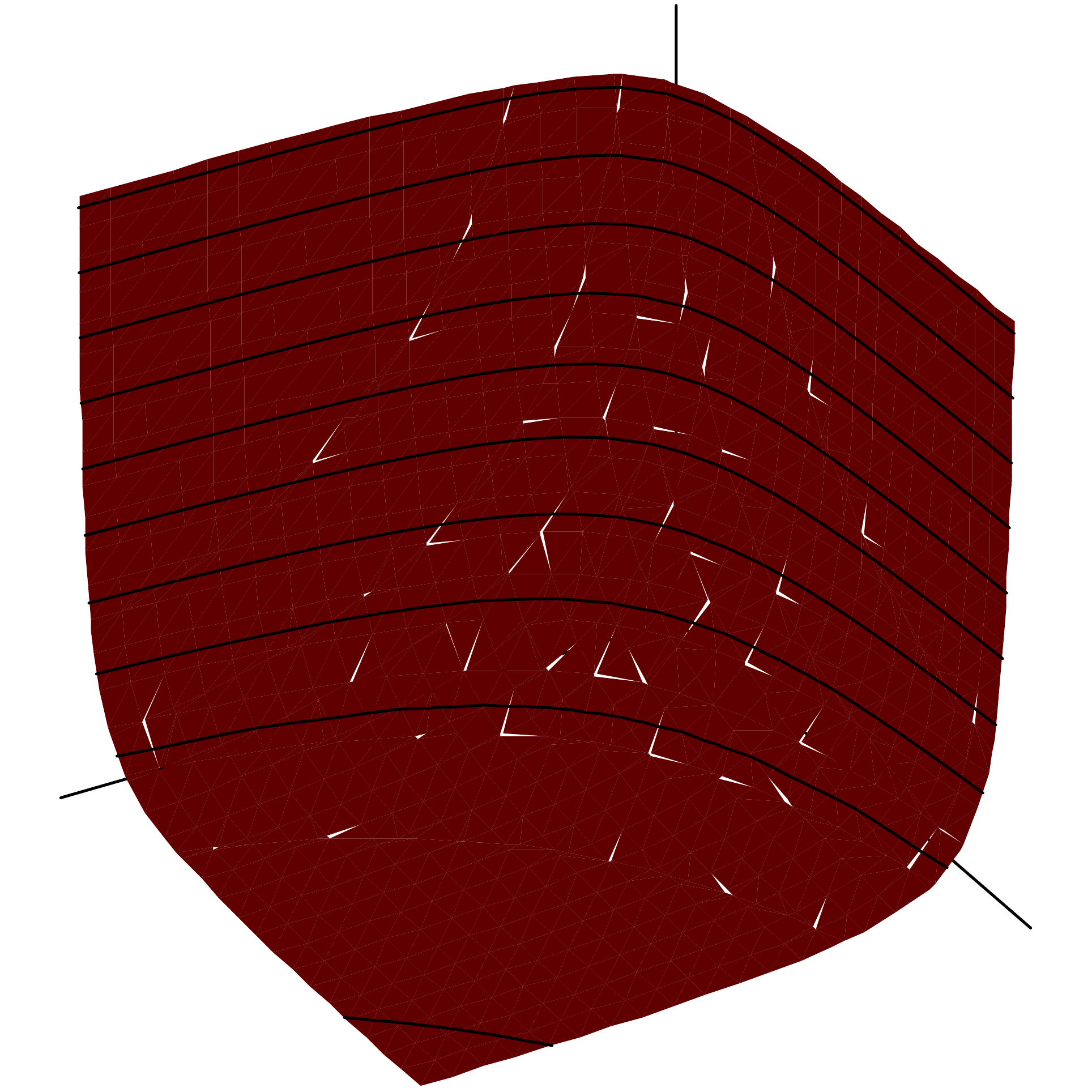}
	\caption{Surface of constant $B_{u}$ in $\{y_u, y_c, y_t\}$ space.}
	\label{fig3}
\end{center}
\end{minipage}
\end{figure}

In view of these surfaces and the expressions of the invariants, the vevs of the fields shall satisfy the hierarchy:

\begin{equation}
\frac{\left\langle B_u\right\rangle}{\Lambda_f^3}\ll\frac{\left\langle A'_{uu}\right\rangle}{\Lambda_f^4}\ll\frac{\left\langle A_u\right\rangle}{\Lambda_f^2}\,.
\label{hierarU}
\end{equation}

The same analysis for the down sector leads to the following relation:
\begin{equation}
\frac{\left\langle B_d\right\rangle}{\Lambda_f^3}\sim\frac{\left\langle A'_{dd}\right\rangle}{\Lambda_f^4}\ll\frac{\left\langle A_d\right\rangle}{\Lambda_f^2}\,.
\label{hierarD}
\end{equation}

\begin{minipage}{0.6\textwidth}
The geometrical analysis allows to interpret  the vevs of the invariant operators as geometric quantities, assuming the hierarchy in eqs. (\ref{hierarU}), (\ref{hierarD}); as can be seen in the figure:
\begin{enumerate}
\item $\left\langle A_i\right\rangle$ sets the radius of the sphere, therefore sets the value of the highest mass.
\item The value of $\left\langle A'_{ii}\right\rangle$ determines how close is the surface in figure \ref{fig2} to the axis. The intersection of this curve and the sphere is a circle around the axis, and the radius of such circle is related to the second highest value of mass.
\item The quantity $\left\langle B_i\right\rangle$ sets the distance of the surface shown in figure \ref{fig3} to the planes $y_t\,y_c$,  $y_c\,y_u$ and  $y_u\,y_t$. This surface, considered in the plane of the circle determined by the intersection of the previous surfaces, is an hyperbola so that the graphic image connects to that for the case of two families.
\end{enumerate}
\end{minipage}
\begin{minipage}{0.35\textwidth}
\begin{center}
	\includegraphics[width=\textwidth]{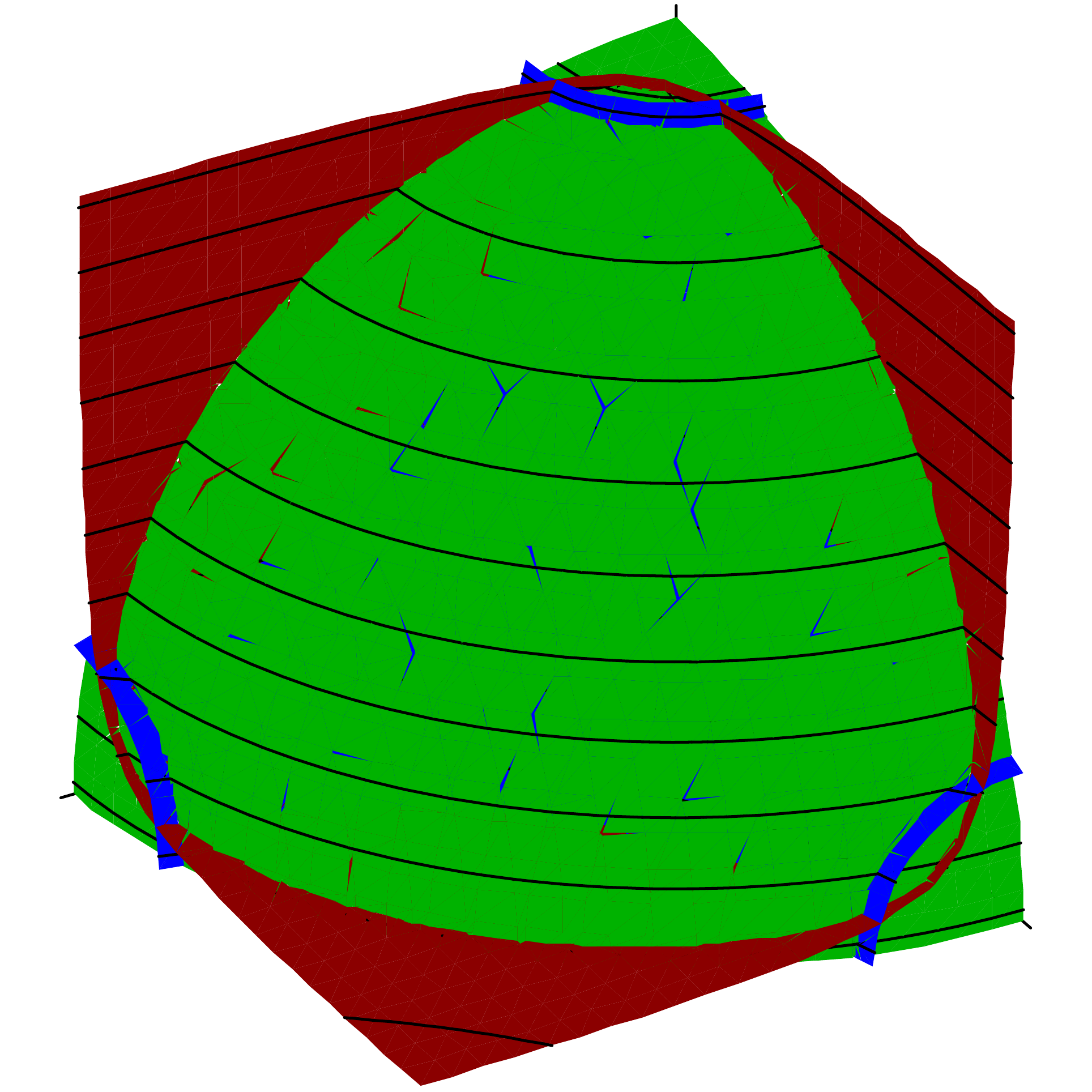}
	Determination of the minimum for a positive definite potential constructed with the invariants in eq. (\ref{inv3fam}).
\end{center}
\end{minipage}\\

The requirements of eq.~(\ref{hierarU}) are not naturally obtained from a general potential, the typical ansatz 
to fix the vev of the invariants being through a ``double-well'' potential of the type:
\begin{equation}
 V_{f-t}=\lambda_u\left(A_u-v_{A_u}\right)^2+\frac{\gamma_u}{\Lambda_{fl}^3}\left(B_u-v_{B_u}\right)^2
 +\frac{\gamma'_u}{\Lambda_{fl}^4}\left(A_{uu}'-v_{A_{uu}'}\right)^2 \,.
\end{equation}
However, for writing this kind of potential, one has to neglect many cross terms that would typically spoil the hierachy. 
Again, the argument proposed here only illustrates a possible, clearly not natural, way to fix the quark masses.
 
The mixing angles appear in the potential through the operator $A_{ud}$:
 \begin{equation}
 \begin{array}{l}
V^{(4)}\supset\lambda_{ud}A_{ud}=\lambda_{ud}\left(P_0+P_{int}\right)\,,\\[3mm]
P_0=-\sum_{i<j} \left(y_{u_i}^2-y_{u_j}^2 \right)  \left( y_{d_i}^2-y_{d_j}^2 \right) \sin^2 \theta_{ij}\,,\\[3mm]
\begin{split}
P_{int}=&\sum_{i<j,k}\left(y_{d_i}^2-y_{d_k}^2 \right)  \left( y_{u_j}^2-y_{u_k}^2 \right) \sin^2 \theta_{ik}\sin^2 \theta_{jk}+\\
&-\left( y_d^2-y_s^2 \right)\left( y_c^2-y_t^2 \right) \sin^2\theta_{12} \sin^2\theta_{13} \sin^2\theta_{23}+ \\
&+\dfrac{1}{2}\left( y_d^2 -y_s^2\right)\left( y_c^2-y_t^2 \right)\cos\delta\, \sin2\theta_{12}\sin2\theta_{23}\sin\theta_{13}\,.
\end{split} 
\end{array}
\end{equation} 
Neglecting the Yukawa couplings for the first family, the equations determining the angles at the minimum of the potential are given by
\begin{equation}
\begin{array}{l}
c_{12}c_{23}s_{12}s_{23}s_{13}\sin\delta=0\,,\\[2mm]
s_{12}c_{12}\left[y_c^2+y_t^2\left(s_{23}^2-s_{13}^2c_{23}^2\right)\right]-y_t^2s_{13}s_{23}c_{23}\left(c^2_{12}-s^2_{12}\right)\cos{\delta}=0\,,\\[2mm]
s_{23}c_{23}\left[y_b^2-y_b^2s_{13}^2+y_s^2s^2_{12}\left(1+s^2_{13}\right)\right]-y_s^2s_{13}s_{12}c_{12}\left(c^2_{23}-s^2_{23}\right)\cos{\delta}=0\,,\\[2mm]
s_{13}c_{13}\left(1-s_{23}^2\right)\left(y_b^2-y_s^2s^2_{12}\right)-y_s^2s_{12}c_{12}s_{23}c_{23}c_{13}\cos{\delta}=0\,,
 \end{array}
 \end{equation}
where $c_{ij}$ and $s_{ij}$ stand for $\cos\theta_{ij}$ and $\sin\theta_{ij}$. The last three equations can be combined into:
\begin{eqnarray}
&&s^2_{12}c^2_{12}\left[y_c^2+y_t^2\left(s_{23}^2-s_{13}^2c_{23}^2\right)\right]^2=y_t^4s_{13}^2s_{23}^2c_{23}^2(1-2s_{12}^2)^2\cos^2{\delta}
\label{eqangle12}\,,\\
&&s_{23}^2\left(y_b^2+y_b^2s_{13}^2+y_s^2 s_{12}^2c_{13}^2\right)=s_{13}^2(y_b^2-y_s^2 s_{12}^2)
\label{eqangle23}\,,\\[1mm]
 &&s_{13}^2c^2_{13}c^2_{23}\left(y_b^2-y_s^2 s^2_{12}\right)\left[y_b^2-y_s^2\left(\cos^2{\delta}s_{12}^2+\sin^2{\delta}s^4_{12}\right)\right]=0\,.
 \label{eqangle13}
 \end{eqnarray}
From eq. (\ref{eqangle13}) it follows that $\sin\theta_{13}=0$ is a solution. Neglecting this angle, $\sin\theta_{23}=0$ can be derived from eq. (\ref{eqangle23}). Finally, from eq. (\ref{eqangle12}), it would result $\sin\theta_{12}=0$. The other alternatives: $\cos\theta_{13}=0$ or $\cos\theta_{13}=0$ lead to unphysiscal solutions but stand as nonvanishing angle configurations and therefore  a novel -if unrealistic- possibility with respect to the two family case.

 %
 %

%
%
%
%
%

 \mathversion{bold}
 \appendixC{C \hspace{0.2cm} The scalar potential for the fundamental approach}
 \mathversion{normal}

There are five independent invariant operators that can be constructed with four fields in fundamental representations of the flavour group, as shown in eqs. (\ref{listfundamentals}) and (\ref{listfundamentals3d}). These invariant operators can be arranged in a vector: denoting this vector by $X^2$ and by $\mean{X^2}$ its vev,
\begin{equation}
\begin{gathered}
X^2\equiv\left(\chi_u^{L\dagger}\chi_u^L\,\,,\,\chi_d^{L\dagger}\chi_d^L\,\,,\,\chi_u^{R\dagger}\chi_u^R\,\,,\,\chi_d^{R\dagger}\chi_d^R\,\,,\,\chi_d^{L\dagger}\chi_u^L\right)^T\,,\\
\left\langle X^2\right\rangle \equiv \left(\left|\chi_u^L\right|^2\,\,,\,\left|\chi_d^L\right|^2\,\,,\,\left|\chi_u^R\right|^2\,\,,\,\left|\chi_d^R\right|^2\,\,,\,\left\langle\chi_u^{L\dagger}\chi_d^L\right\rangle\right)^T\,.
\end{gathered}
\end{equation}

All these invariant operators have dimension two and the most general renormalizable scalar potential is given by:
\begin{equation}
V_\chi=-\frac{1}{2}\sum_i\left(\mu_i^2X_i^2+h.c.\right)+\sum_{i,j}\lambda_{ij}\left(X_i^2\right)^*X_j^2=
-\frac{1}{2}\left[\mu^2X^2+h.c.\right]+\left(X^2\right)^\dagger\lambda\,X^2\,,
\end{equation}
where $\lambda$ is a $5\times5$ hermitian matrix\footnote{indices run over the five values \{$uL$,$dL$,$uR$,$dR$,$ud$\}.} 
and the mass terms are arranged in the vector $\mu^2$.  There are therefore a total of 20 invariant operators in the 
most general renomalizable potential.
Assuming that $\lambda$ is invertible and adding a constant term to the potential the above expression can be rewritten as:
\begin{equation}
V_{\chi}=\left(X^2-\dfrac{1}{2}\lambda^{-1}\,\mu^2\right)^\dagger\lambda\left(X^2-\dfrac{1}{2}\lambda^{-1}\,\mu^2\right)\,.
\end{equation}
For a bounded-from-below potential, $\lambda$ has to be positive definite which implies that the minimum of the scalar potential is reached for:
\begin{equation}
\left\langle X^2\right\rangle=\frac{1}{2}\lambda^{-1}\mu^2\,.
\label{solSfund}
\end{equation}
This is the formal expression for the minimum. Yukawa eigenvalues and the Cabibbo angle are related to the configuration of the potential minimum through eq.~(\ref{connectionfund}), which together with the previous equation yield~\footnote{$\left(\lambda^{-1}\mu^2\right)_{i}=\sum_j\left(\lambda^{-1}\right)_{ij}\mu^2_j$}:
\begin{equation}
\begin{gathered}
y_c^2=\frac{1}{4\Lambda_f^4}\left(\lambda^{-1}\mu^2\right)_{uL}\left(\lambda^{-1}\mu^2\right)_{uR}\,,\qquad \,
y_s^2=\frac{1}{4\Lambda_f^4}\left(\lambda^{-1}\mu^2\right)_{dL}\left(\lambda^{-1}\mu^2\right)_{dR}\,,\\
\cos\theta_c=\frac{\left(\lambda^{-1}\mu^2\right)_{ud}}{\sqrt{\left(\lambda^{-1}\mu^2\right)_{dL}\left(\lambda^{-1}\mu^2\right)_{uL}}}	\,.
\end{gathered}
\end{equation}
 Remarkably, naturalness criteria imply $\cos\theta_c\sim \cO(1)$ at this very general level. Yukawa eigenvalues are $\cO(\mu^2/\lambda\Lambda_f^2)$, implying $\sqrt{(\lambda^{-1}\mu^2)_{uR,uL}}\sim 10^{-1}\Lambda_f$ in order to fix the charm Yukawa eigenvalue to the observed value and $\sqrt{(\lambda^{-1}\mu^2)_{dR,dL}}\sim 10^{-2}\Lambda_f$ to fix analogously the strange Yukawa eigenvalue.\\

For the sake of clarity and definiteness, we present next an example of a scalar potential whose mass parameters are directly connected to the Yukawa couplings. We assume that RH flavons acquire vevs equal to $\Lambda_f$, then the parametrization in eq.~(\ref{vevsFundamentals2D}) follows. Such assumption can be justified by naturalness arguments or simply fixing the parameters associated to $\left|\chi^R_{d,u}\right|$ through eq.~(\ref{solSfund}). We can then concentrate only on the scalar potential for the LH flavons:
\begin{equation}
V^\prime_\chi=\lambda_u\left(\chi_u^{L\dagger}\chi_u^L-\frac{\mu_u^2}{2\lambda_u}\right)^2+\lambda_d\left(\chi_d^{L\dagger}\chi_d^L-\frac{{\mu}_d^2}{2\lambda_d}\right)^2
 +\lambda_{ud}\left(\chi_u^{L\dagger}\chi_d^L-\frac{\mu_{ud}^2}{2\lambda_{ud}} \right)^2\,.
\end{equation}
As already stated, at the minimum the invariants in this potential can be written in terms of Yukawa eigenvalues and the Cabibbo angle:
\begin{equation}
V^\prime_\chi=\lambda_u\left(\Lambda_f^2y_c^2-\frac{\mu_u^2}{2\lambda_u}\right)^2+\lambda_d\left(\Lambda_f^2y_s^2-\frac{{\mu}_d^2}{2\lambda_d}\right)^2
 +\lambda_{ud}\left(\Lambda_f^2y_cy_s\cos{\theta_c}-\frac{\mu_{ud}^2}{2\lambda_{ud}} \right)^2\,.
\end{equation}
From this relation, the expression for Yukawa eigenvalues and the Cabibbo angle in terms of the parameters of the potential can be read:
\begin{equation}
y_c=\frac{\mu_u}{\sqrt{2\lambda_{u}}\Lambda_f}\,,\qquad\qquad 
y_s=\frac{\mu_d}{\sqrt{2\lambda_{d}}\Lambda_f}\,,\qquad\qquad
\cos\theta_c=\frac{\sqrt{\lambda_{u}\lambda_{d}}\mu_{ud}^2}{\lambda_{ud}\,\mu_d\,\mu_u}\,.
\end{equation}
The resulting $\cos\theta_c$ is naturally of $\cO(1)$, while correct charm and strange masses arise when $\mu_u\sim10^{-2}\sqrt{\lambda_u}\Lambda_f$ and $\mu_d\sim10^{-3}\sqrt{\lambda_d}\Lambda_f$. The differences with the bi-fundamental approach can be seen comparing the above equation with eq.~(\ref{solbifund}).  The example shown corresponds to a  potential with some terms omitted\footnote{Terms like $g_{ud}\chi_d^{L\dagger}\chi_d^L\chi_u^{L\dagger}\chi_u^L$ do not affect the position of the minimum provided $g_{ud}<y_s^2/y_c^2$.}, whose mere purpose is to illustrate explicitly the mechanism of generation of Yukawa eigenvalues and mixing angle through a scalar potential for the $d=6$ Yukawa operator.

Finally, notice that the extension to the three family case is trivial, substituting in the formulae above $y_c$ and $y_s$ by $y_t$ and $y_b$, respectively, and the Cabibbo angle by $\theta_{23}$. This stems from the fact that, considering the renormalizable scalar potential, only the heaviest Yukawas are non-vanishing, as discussed in the main text.

%
%

\bibliographystyle{plain}

\begin{thebibliography}{10}
\bibliographystyle{plain}


\bibitem{DGIS:MFV}
G.~D'Ambrosio, G.~Giudice, G.~Isidori, and A.~Strumia, Nucl. Phys. {\bf B645}
  {(2002)} 155--187 {[arXiv:hep-ph/0207036]}.

\bibitem{PQ:U1PC}
R.~D. Peccei and H.~R. Quinn, Phys. Rev. {\bf D16} {(1977)} 1791--1797.

\bibitem{Buras:MFVandBeyond}
For a recent review on the subject, see:
A.~J. Buras, arXiv: 1012.1447.

\bibitem{Branco}
 G.C. Branco, W. Grimus and L. Lavoura, Phys. Lett. {\bf B380} {(1996)} 119 {[arXiv:hep-ph/9601383]}.
 
\bibitem{Kagan}
A. L. Kagan et al., Phys. Rev. {\bf D80} {(2009)} 076002 {[arXiv:0903.1794]}.

\bibitem{INP:GenericFS}
  G.~Isidori, Y.~Nir, G.~Perez, arXiv:1002.0900.

  \bibitem{Isidori2-bis} 
  G. Isidori, arXiv:1012.1981, and references therein.

\bibitem{LPR:MFVsusy}
Z.~Lalak, S.~Pokorski, and G.~G. Ross, JHEP {\bf 08} {(2010)} 129 {[arXiv:1006.2375]}.

\bibitem{FPR:MFVextraD}
A.~Fitzpatrick, G.~Perez, and L.~Randall, [arXiv:0710.1869].

\bibitem{CG:MFV}
R.~S. Chivukula and H.~Georgi, Phys. Lett. {\bf B188} {(1987)} 99.

\bibitem{Syms}
  C.~D.~Froggatt, H.~B.~Nielsen,
  Nucl.\ Phys.\  {\bf B147 } (1979)  277;\\
  For a recent review on the subject, see:
  G.~Altarelli, F.~Feruglio,
  Rev.\ Mod.\ Phys.\  {\bf 82 } (2010)  2701-2729.
  [arXiv:1002.0211].
  
\bibitem{GRV:SU3gauged}
B.~Grinstein, M.~Redi, and G.~Villadoro, JHEP {\bf 11} {(2010)} 067 {[arXiv:1009.2049]}.

\bibitem{Feldmann:SU5gauged}
T.~Feldmann, arXiv:1010.2116.

\bibitem{CGIW:MLFV}
V.~Cirigliano, B.~Grinstein, G.~Isidori, and M.~B. Wise, Nucl. Phys. {\bf B728}
  {(2005)} 121--134 {[arXiv:hep-ph/0507001]}.
  
\bibitem{DP:MLFV}
  S.~Davidson, F.~Palorini,
  Phys.\ Lett.\  {\bf B642 } (2006)  72--80 [arXiv:hep-ph/0607329].

\bibitem{GHHH:MFSM}
M.~B. Gavela, T.~Hambye, D.~Hernandez, and P.~Hernandez, JHEP {\bf 09} {(2009)}
  038 {[arXiv:0906.1461]}.
  
 \bibitem{PDG2010}
K.~Nakamura {\it et al.} [ Particle Data Group Collaboration ],
  J.\ Phys.\ G {\bf G37 } (2010)  075021.
  
\bibitem{BGI:MFVwithRHcurrents}
T.~Feldmann, T.~Mannel,
  JHEP {\bf 0702 } (2007)  067.
  [hep-ph/0611095];
A.~J. Buras, K.~Gemmler, and G.~Isidori, Nucl. Phys. {\bf B843} {(2011)}
  107--142 {[arXiv:1007.1993]}.
  
\bibitem{Feldmann:2009dc}
  T.~Feldmann, M.~Jung, T.~Mannel,
  Phys.\ Rev.\  {\bf D80 } (2009)  033003.
  [arXiv:0906.1523 [hep-ph]].

\bibitem{CNS:SUSYMFVRunning}
G.~Colangelo, E.~Nikolidakis, and C.~Smith, Eur. Phys. J. {\bf C59} {(2009)}
  75--98 {[arXiv:0807.0801]}.

\bibitem{MS:SUSYMFVEDM}
L.~Mercolli and C.~Smith, Nucl. Phys. {\bf B817} {(2009)} 1--24 {[arXiv:0902.1949]}.

\bibitem{AGMR:CPpaper}
R.~Alonso, M.B.~Gavela, L.~Merlo and S.~Rigolin, in preparation.

\bibitem{Berezhiani:2005tp}
  Z.~Berezhiani and F.~Nesti,
  JHEP {\bf 0603} (2006) 041
  [arXiv:hep-ph/0510011].
  
\bibitem{FKR:AnarchicalMasses}
L.~Ferretti, S.~F. King, and A.~Romanino, JHEP {\bf 11} {(2006)} 078 {[arXiv:hep-ph/0609047]}.

\bibitem{CFRZ:UnifiedAnarchicalMasses}
L.~Calibbi, L.~Ferretti, A.~Romanino, and R.~Ziegler, JHEP {\bf 03} {(2009)} 031 {[arXiv: 0812.0087]}.

\bibitem{Nardi:2011st}
  E.~Nardi, [arXiv:1105.1770 [hep-ph]].

\bibitem{Planck2011}
  B.~Gavela, plenary talk at Planck 2011, 30th  of May - 3rd of June 2011, Lisbon.
\end{thebibliography}

\end{document}